\documentclass[nonatbib, final,5p,times,twocolumn]{elsarticle}

\makeatletter
\let\c@author\relax

\makeatother
\newcommand{\thesisTitle}{The Value of Competing Energy Storage in Decarbonized Power Systems}
\newcommand{\thesisName}{Maximilian Parzen}
\newcommand{\thesisSubject}{Reserach paper}

\PassOptionsToPackage{% setup clean thesis style
    figuresep=colon,%
    hangfigurecaption=false,%
    hangsection=true,%
    hangsubsection=true,%
    sansserif=false,%
    configurelistings=true,%
    colorize=full,%
    colortheme=bluemagenta,%
    configurebiblatex=true,%
    bibsys=bibtex,%
    bibfile=references,%
    bibstyle=numeric,%
    bibsorting=none,%
}{cleanthesis}
\usepackage{cleanthesis}

\hypersetup{% setup the hyperref-package options
    pdftitle={\thesisTitle},    %   - title (PDF meta)
    pdfsubject={\thesisSubject},%   - subject (PDF meta)
    pdfauthor={\thesisName},    %   - author (PDF meta)
    plainpages=false,           %   -
    colorlinks=true,
    linkcolor=ctcolorparttext,
    filecolor=ctcolortitle,      
    urlcolor=ctcolortitle,
    citecolor=ctcolortitle,
    pdfborder={0 0 0},          %   -
    breaklinks=true,            %   - allow line break inside links
    bookmarksnumbered=true,     %
    bookmarksopen=true          %
}

\journal{arXiv. Not peer-reviewed}

\begin{document}

\begin{frontmatter}

\title{The Value of Competing Energy Storage in Decarbonized Power Systems}

\author[First, Second]{Maximilian Parzen}
\ead{max.parzen@ed.ac.uk}
\author[Third]{Davide Fioriti}
\author[First]{Aristides Kiprakis}
\address[First]{University of Edinburgh, Institute for Energy Systems, EH9 3DW Edinburgh, United Kingdom}
\address[Third]{University of Pisa, Department of Energy, Systems, Territory and Construction Engineering, Largo Lucio Lazzarino, 56122 Pisa, Italy}
% \address[Second]{Open Energy Transition, Koenigsallee 52, 95448 Bayreuth, Germany}

\begin{abstract}
As the world seeks to transition to a sustainable energy future, energy storage technologies are increasingly recognized as critical enablers. However, the macro-energy system assessment of energy storage has often focused on isolated storage technologies and neglected competition between them, thus leaving out which energy storage to prioritise. The article applies a systematic deployment analysis method that enables system-value evaluation in perfect competitive markets and demonstrates its application to 20 different energy storage technologies across 40 distinct scenarios for a representative future power system in Africa. Here, each storage solution is explored alone and in competition with others, examining specific total system costs, deployment configuration, and cost synergies between the storage technologies. The results demonstrate the significant benefits of optimizing energy storage with competition compared to without (+10\% cost savings), and highlight the relevance of several energy storage technologies in different scenarios. This work provides insights into the role of energy storage in decarbonizing power systems and informs future research and policy decisions. There is no one-size-fits-all energy storage, but rather an ideal combination of multiple energy storage options that are designed and operated in symbiosis.
\end{abstract}


\end{frontmatter}

% Abbreviations
\begin{acronym}[UMLX]
    \acro{BAU}{Business as Usual}
    \acro{EP}{Energy to Power}
    \acro{LCOS}{Levelized Cost of Storage}
    \acro{LCOE}{Levelized Cost of Electricity}
\end{acronym}

%%%%%%%%%%%%%%%%%%%%%
%%%%%%%%%%%%%%%%%%%%%
\section{Introduction}\label{mppafrica:sec:introduction}

As the world looks to decarbonise its power systems in order to mitigate the impacts of climate change, power modeling scenarios have made it increasingly clear that energy storage will play a critical role in the transition to a more sustainable future \cite{Sepulveda2021TheSystems, Albertus2020Long-DurationTechnologies, Arbabzadeh2019TheProduction, Mallapragada2020Long-runGeneration, Tong2020}. The rise of renewable energy sources such as solar and wind power has presented a significant challenge for the electricity grid, which must balance the variable and intermittent nature of these sources with the electricity demand. Energy storage technologies provide a solution to this challenge by allowing excess renewable energy to be stored and used when needed, effectively decoupling the generation and consumption of electricity while adding system-value. Here, as in \cite{Umamaheswaran2015}, the system-value of energy storage refers to the broader economic benefits that storage can provide to the power system beyond its immediate application. These benefits include the displacement of firm generation and network infrastructure, greater renewable energy utilisation, and the reduction of transmission and distribution losses, which often reduces the reliance on fossil fuels and lowers carbon emissions. In this context, energy storage, with its system-value provision, is a key enabler of transitioning to a cleaner, more sustainable energy system worldwide.

According to \cite{Parzen2021BeyondSystems}, assessing the competitiveness or suitability of energy storage in larger power systems with well-known \ac{LCOS} methods as applied in \cite{Schmidt2019, Beuse2020ProjectingSector, Tafone2020LevelisedCycle, Julch2016ComparisonMethod} are less suitable compared to system-value assessment methods as applied \cite{deSisternes2016TheSector, Sepulveda2021TheSystems, Mallapragada2020Long-runGeneration, Georgiou2020, Ebbe2022}. However, all these system-value assessments explore isolated storage technologies that do not consider any competition with other storage technologies.
For instance, the inspiring work in \cite{Sepulveda2021TheSystems} assessed a single generic energy storage in two representative decarbonised power systems.
Through a design-space exploration of the generic storage, they identified what energy capacity costs are required to replace firm generation, which are the most critical storage performance parameters and sizing characteristics that contribute to the system-value. 
However, it is also known that adding more technology options to models often results in synergies. These synergies reduce the significantly total system costs, which are defined as the sum of all operational and investment costs, raising at least questions of the validity of the previously found results of single energy storage scenarios \cite{Parzen2021BeyondSystems}.
%What likely prohibited the inclusion of competition traditional studies is the reliance on counterfactual studies, which compare one base case e.g. without storage to one with storage which comes not handy when trying to attribute the system value to multiple technologies at the same time. 
%Evolving from these insights, \citet{Parzen2021BeyondSystems} describes that competition has been probably ignored in this type of analysis because the methods were not suitable.
Expanding on this knowledge, \cite{Parzen2021BeyondSystems} introduces and demonstrates a systematic deployment analysis method that enables system-value evaluation in perfect competitive markets but demonstrates the method by ignoring uncertainty and only considering a limited amount of storage technologies, namely hydrogen and lithium-ion energy storage. 

\begin{figure*}[ht]
\centering
%\hspace{-25pt}
\includegraphics[trim={3.3cm 11cm 3.1cm 6cm} ,clip,width=\textwidth]{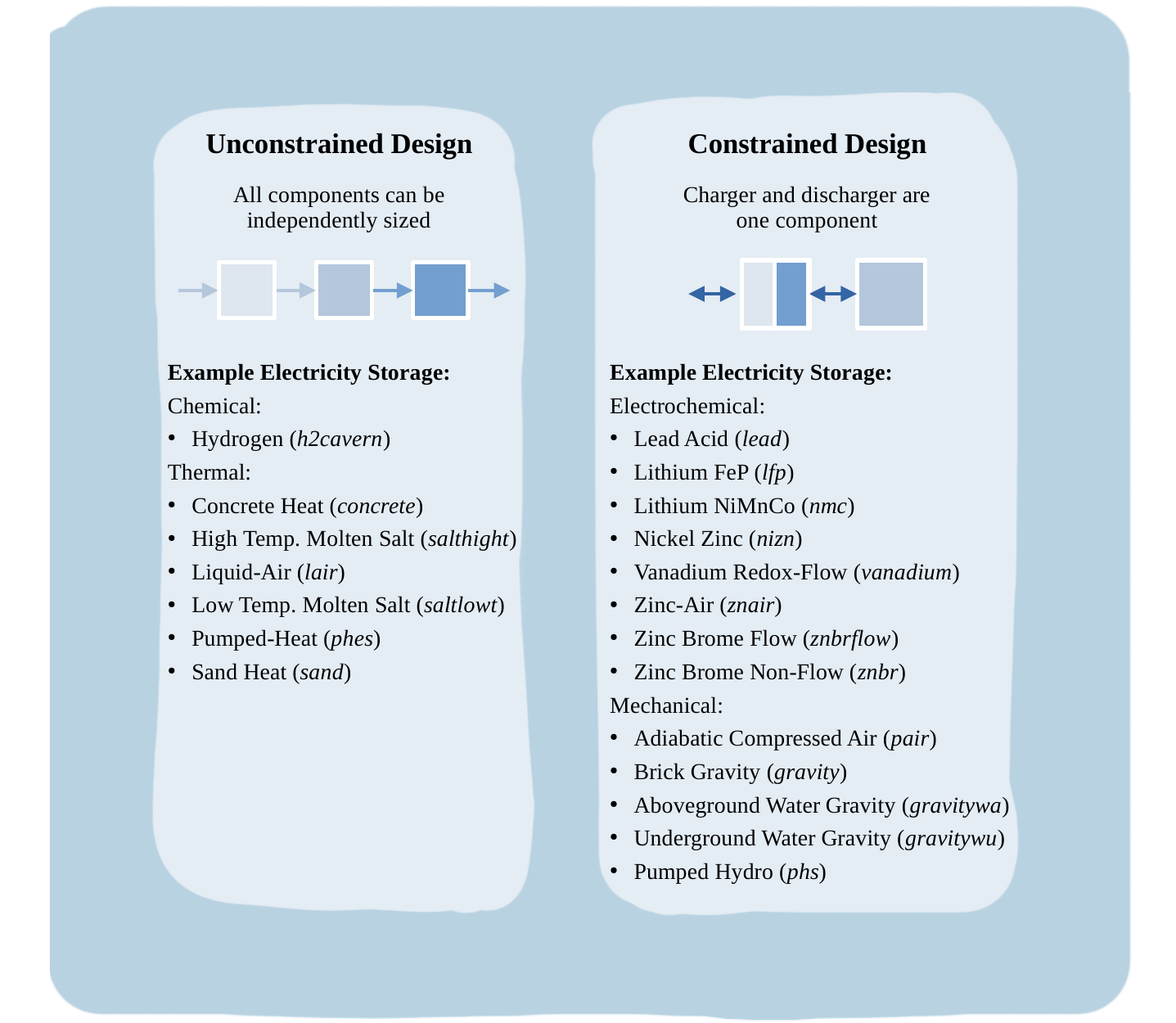}
\caption{Illustration of energy storage technologies with abbreviations used in this study.}
\label{aueu-fig:foundation:storage-classification}
\end{figure*}

In this article, we assess multiple energy storage with the newly suggested systematic deployment analysis, also addressing uncertainty. In total, we assess the system-value of 20 energy storage (see Figure \ref{aueu-fig:foundation:storage-classification}) with and without competition across 40 distinct scenarios for a representative future power system in Africa.
We use a global coverage open energy system model suitable for investment and operational co-optimization, including grid infrastructure and detailed operating decisions and constraints \cite{Parzen2022PyPSAEarth}.
Further, we apply this model to its already validated Nigerian power system \cite{Parzen2022PyPSAEarth}, configure it with high temporal resolution (8760h) and a spatial interconnected 10-node system to keep some of the underlying grid and environmental information within the simplification.
Within this model, we integrate for the first time 20 storage technologies, which data is collected and expanded from Pacific Northwest National Laboratory (PNNL) (see Table \ref{aueu-tab:2050-storage-assumptions}, and Methods \ref{aueu:sec:methods:datacollection}).
We explore two unanswered questions: how significant is the system-benefit from optimizing energy storage with competition compared to without, and which energy storage is optimization relevant considering uncertainty. 

\begin{figure*}[ht]
\centering
%\hspace{-25pt}
\includegraphics[trim={0cm 0cm 0cm 0cm},clip,width=0.8\textwidth]{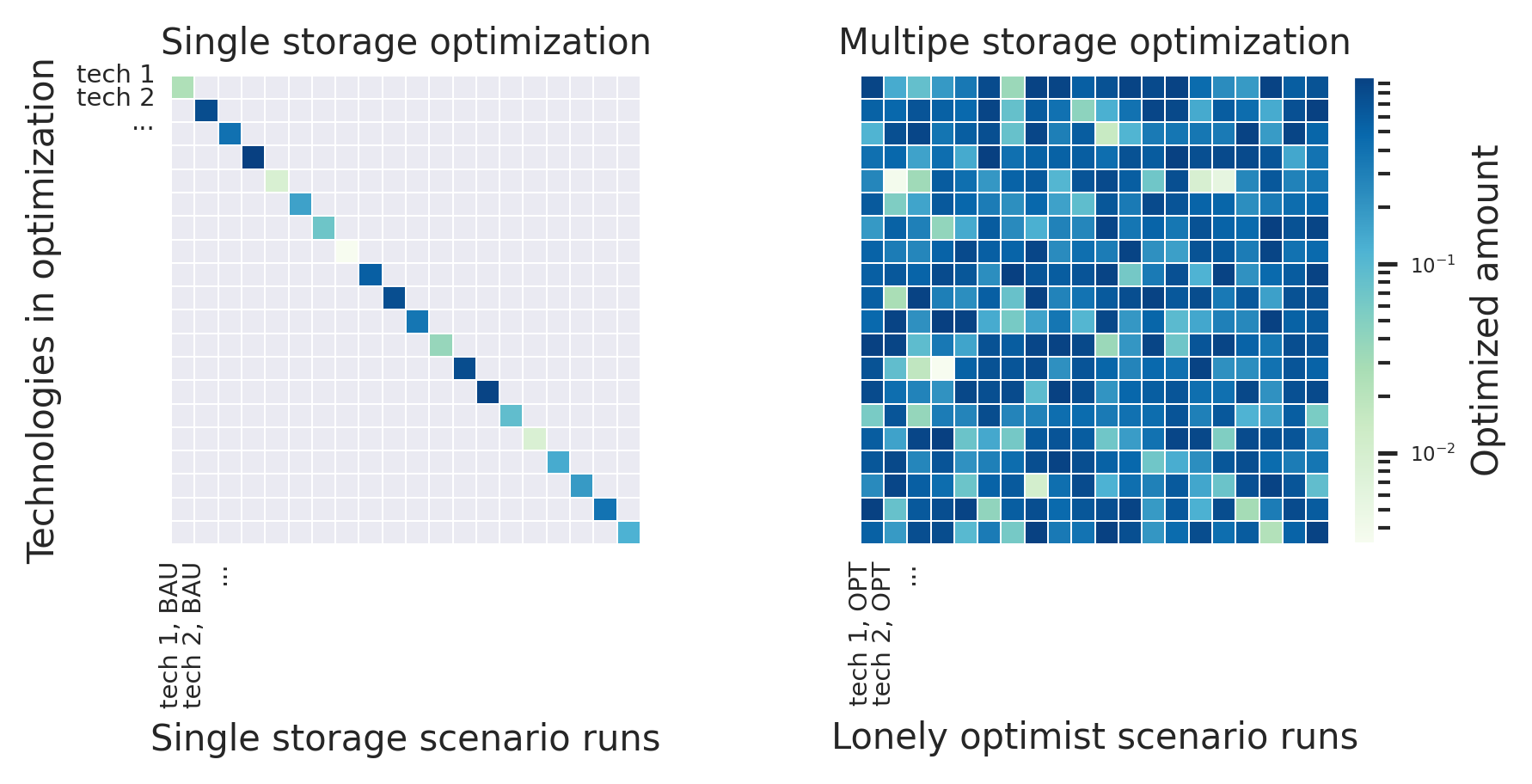}
\caption{Illustration of scenario concept in this study.}
\label{aueu-fig:scenario-set-demo}
\end{figure*}

To answer the research questions, we focused on two scenario trees as illustrated in Figure \ref{aueu-fig:scenario-set-demo}.
The first scenario tree, defined as 'single storage' scenario, involves optimizing each of the energy storage solutions in isolation, assuming business-as-usual costs. This scenario set includes 20 optimization runs and excludes any competition between the different storage solutions. By doing so, it is possible to investigate the specific total system costs (\EUR{}/MWh) and deployment configuration (GW for charger/discharger or GWh for store).
The second scenario tree also involves 20 optimization runs, but in this case, all energy storage solutions can be optimized within each scenario. This approach allows for perfect competition and cost synergies between the different technologies. To facilitate this, this article uses the here coined 'lonely optimist' approach, where one storage option has optimistic capital costs while the others have pessimistic assumptions. This extreme parameterization enables us to suggest which energy storage solutions provide system-value and which can potentially be neglected - at least within the modelled power system conditions.
%The final 'plan B' scenario tree with 10 optimization runs, optimizes first all technologies under the business-as-usual cost assumption and excludes from all following scenarios the most deployed store and charger unit until no energy storage is left. 
By applying the new systematic deployment analysis from \cite{Parzen2021BeyondSystems}, this manuscript suggests that optimizing scenarios with multiple energy storage, compared to scenarios with single energy storage, can lead to significant system benefits between $3-29\%$. Considering the extreme parameterization, it was also found that 9 out of 20 storage technologies are optimization-relevant, often providing system benefits due to synergies in storage design and operation. The often praised Lithium energy storage \cite{Schmidt2019, Beuse2020ProjectingSector} was only found highly competitive in a single scenario with optimistic cost assumption of (\EUR{112}/kWh and \EUR{24}/kW). In contrast, the often studied hydrogen storage was indeed consistently optimization relevant as well as the sand-based thermal energy storage. However, other technologies added competitive pressure, including gravity-brick, underground and above-ground water-based gravity and pump-heat energy storage, compressed-air and nickel-based electricity storage. Therefore, the system-value technology assessments with multiple energy storage technologies can be considered as an advanced conceptual approach that could find more application in research and industry compared to approaches that ignore competition by isolated technology considerations. 

% The most significant and frequent optimized technologies in the analysed power system include the gravity, hydrogen-cavern, compressed-air and sand-based thermal energy storage indicating their importance to deliver system-value. Surprisingly, the Lithium Ferrous Phosphate (LFP) storage is only highly competitive in a single scenario with optimistic cost assumption of (\EUR{112}/kWh and \EUR{24}/kW). 
% Similar, other energy storage technologies including the Lithium Nickel Manganese Cobalt (NMC) battery, Nickel-Zinc battery, pumped-heat energy storage, can provide relevant synergies to the power system with optimistic cost assumption, however, not to a significant extend. 

% a) how significant is the system-benefit from optimizing energy storage with competition compared to without
% b) which energy storage is optimization relevant considering uncertainty

% CONTRIBUTION:
% - 20 energy storage are optimized in African power system
% - X1, X2, X3, X4 tech are only relevant for optimization
% - Cost uncertainty impact total system cost and optimization amount significant
% - New open-source storage data interface available to any model
% - All storage models directly available in PyPSA-Earth model

% \includegraphics[trim={1.2cm 11cm 1cm 8cm} ,clip,width=\textwidth]

%%%%%%%%%%%%%%%%%%%%
%%%%%%%%%%%%%%%%%%%%
\section{Modelling single vs multiple energy storage}\label{aueu:sec:no-competition}

\begin{figure*}[ht]
\centering
%\hspace{-25pt}
\includegraphics[trim={0cm 0.2cm 0cm 0.2cm},clip,width=0.8\textwidth]{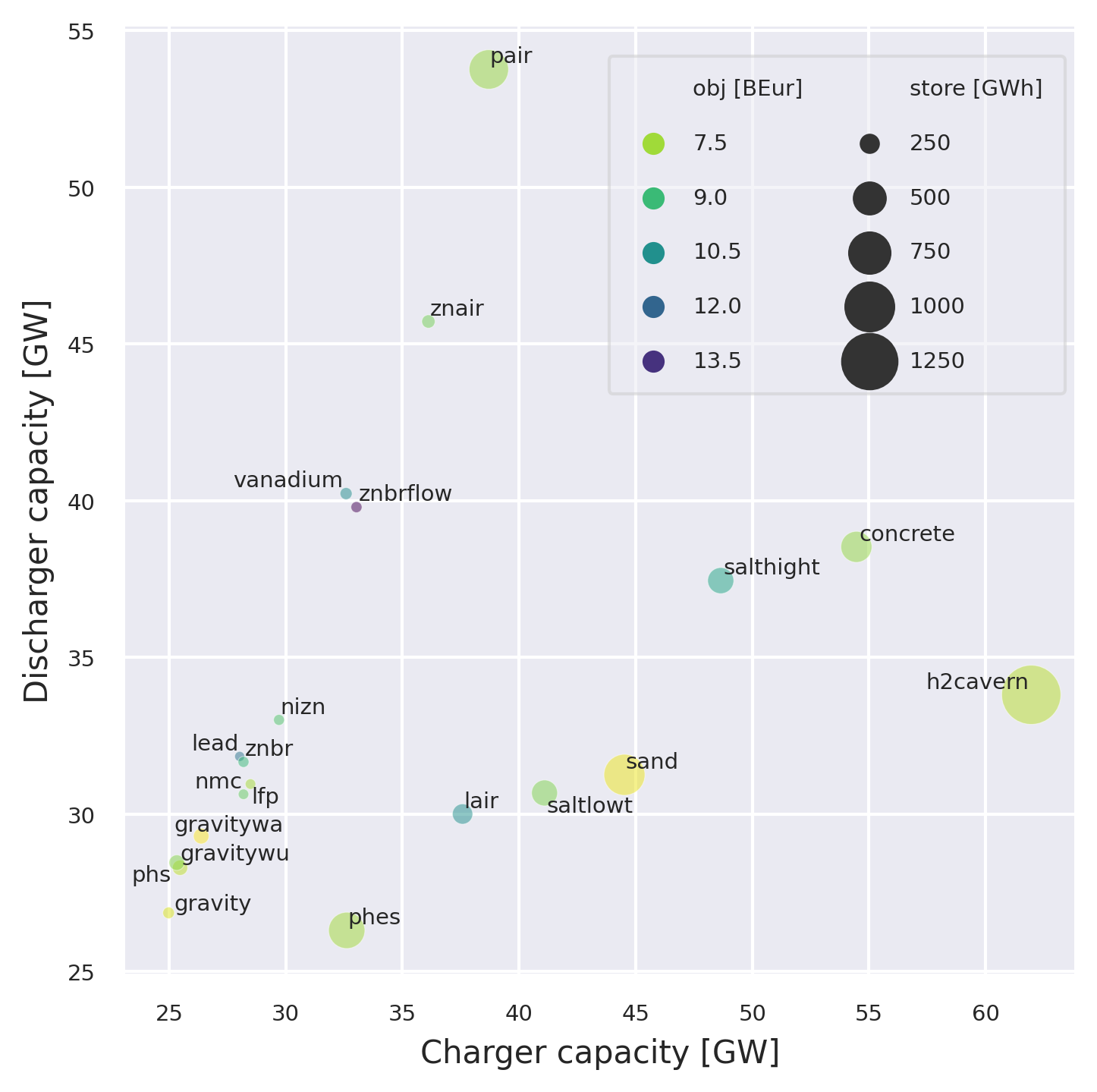}
\caption{Optimization results for single energy storage scenarios. The y-axis, x-axis, and marker size show the deployment required for a least-cost 2050 power system in Nigeria. The colour indicates the total system costs.}\label{aueu-fig:ng-storage-scatterplot}
\end{figure*}

The 'system-value' of technologies can be defined as its market potential resulting from possible and probable least-cost scenarios in capacity expansion models (see Section \ref{aueu:sec:methods}). Figure \ref{aueu-fig:ng-storage-scatterplot} presents a range of optimised market potentials for various single-optimised energy storage technologies. Because only one storage is included in each optimization run, these scenarios represent a case that ignores any competition. It can be observed that the most and least optimised charger technology ranges between $25-54GW$ for the gravity-brick (gravity) and compressed air energy storage (pair), the most and least optimised dischargers ranging between $26-54GW$ for pump-heat (phes) and the compressed air energy storage (pair), and the most and least optimised stores ranging between $0.18-1.46~TWh$ for lead battery (lead) and hydrogen cavern (h2cavern) storage systems, respectively. These results are unrealistic as there are always multiple options available; nevertheless, they reveal that every energy storage technology can serve the energy system or, in other words, contain system-value.

Further, observing Figure \ref{aueu-fig:ng-storage-scatterplot}, one can see that the least-cost system model optimizes various energy storage ratios between charger, store and discharger depending on the technologies. As for most models \cite{Neumann2021TheModel}, storage technologies are constrained such that perfect balancing is guaranteed, ensuring no mismatch between electricity supply and demand. However, there is a general trade-off between storage, grid, and supply expansion, allowing for significantly different storage designs. Theoretically, creating a power system without any storage or grid is feasible if renewables can be massively overbuilt and curtailed. The addition of energy storage to power systems allows for smoothing out mismatches in time, while grid infrastructure helps reduce mismatches in space and to exploit better resource potentials \cite{Hirth2014, Brown2018SynergiesSystem}. Since every storage technology has different capital costs and efficiencies in the component chain (charger, store, discharger), the design of storage technologies changes to exploit its role in the power system to achieve the minimum total system costs.
It is important to note that the scenarios shown in the figure are only one possible outcome of the least-cost power system optimization model, and there may be many other factors that could influence the market potential of a technology. One important factor, making scenarios not only possible but also more probable, is the competition between storage technologies which is discussed next. 

% Therefore, the system-value of a technology should be viewed here as a rough estimate rather than a precise prediction. Nonetheless, this information is valuable for policy makers and investors in making informed decisions about which technologies to invest in for a sustainable and reliable energy system.

% \section{System-value of storage with competition}\label{aueu:sec:no-competition}

\begin{figure*}[ht]
\centering
%\hspace{-25pt}
\includegraphics[trim={0cm 0cm 0cm 0cm},clip,width=0.8\textwidth]{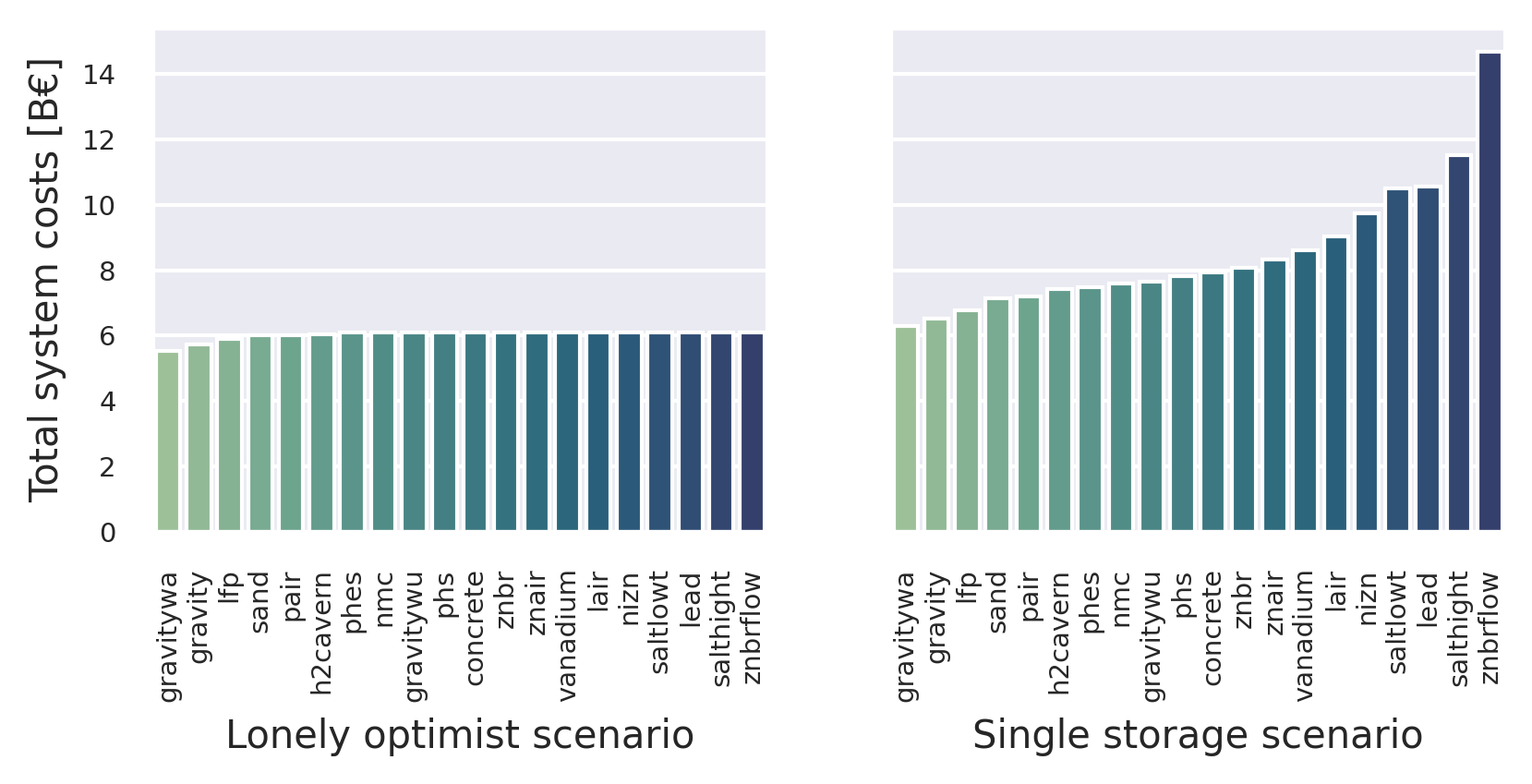}
\caption{Total system cost for energy storage scenario with (left) and without (right) competition. Scenarios are sorted according to the total system costs.}
\label{aueu-fig:competition-vs-nocompetition}
\end{figure*}

Figure \ref{aueu-fig:competition-vs-nocompetition} shows that power systems with perfect storage competition (lonely optimist scenario) are, on average, significantly cheaper compared to those without storage competition (single storage scenario). 
% - Describe flat vs curve. + surprise! Order hasn't changed!
First and most apparent, the competitive scenarios are 29\% cheaper compared to single storage optimization scenarios. While the competitive scenario tree has few cost increases for some technologies initially, the cost gradient becomes relatively low after the optimistic 'phes' scenario with changes of less than $0.1\%$. In contrast, one can observe continuous significant cost increases for single storage scenarios. Surprisingly, comparing both x-axes that are sorted according to total system costs, it was found that the order of cost-optimal storage systems for the power system is identical. This identical order implies that the storage that leads to the lowest total system in the single-optimized storage scenarios, is likely also the most valuable and important storage in the context of contributing to system benefits in the other scenario tree. 
% Evidently large benefit of perfect competing markets
Second and most important, the power systems benefits from synergies provided by a perfect competing storage market even under the worst cost assumptions. The total system of the most expensive storage scenario in the competitive situation is $3\%$ (6.295/6.108) cheaper than the best storage scenario in the non-competitive scenario. This is remarkable because the 'single storage' scenario assumes business-as-usual costs for the most favourable technology, while the most expensive 'lonely optimist' scenario considers  optimistic costs for the least favourable storage technology ($-30\%~\text{of~BAU}$) while simultaneously for all other storage technologies pessimistic costs ($+30\%~\text{of~BAU}$).
Thirdly, comparing the 'gravitywa' storage from both scenario trees, one can find significant cost savings when considering perfect competitive storage markets. When considering 'gravitywa' with \ac{BAU} assumptions in both scenario trees, one can observe $8\%$ cost saving or $500~\text{million~\EUR{}}$ in absolute terms. Interesting, but less of an apple-to-apple comparison, setting the 'gravitywa' technology as optimistic, as given in the original lonely optimist scenario, the savings add up to $13\%$. 

These results suggest that studies that assess the system-value with single optimized energy storage such as \cite{Sepulveda2021TheSystems, Mallapragada2020Long-runGeneration, deSisternes2016TheSector} miss significant benefits from synergies caused by co-optimizing multiple energy storage technologies. It is also likely that power systems with two or three modelled energy storage, such as \cite{Dowling2020RoleSystems, Victoria2019TheSystem}, could benefit from system cost reduction when including more of the technologies that were found highly optimization relevant, for instance, the gravity or sand based thermal energy storage. When assuming similar power system conditions as in the study, the system cost can be up to $5-13\%$ for fully decarbonized power systems.

% - 29\% (6037/8544) lonely optimist vs single storage mean difference
% - 13\% cheaper (5518/6295) gravitywa optimist vs single store BAU
% - 8\% cheaper (5797/6295) gravitywa BAU vs single store BAU
% - 3\% (6.295/6.108) worst competition OPTI vs best single store BAU

\section{Assessing technology importance}\label{aueu:sec:relevancy}

\begin{figure*}[!ht]
% https://stackoverflow.com/questions/52628649/multiple-subfigures-in-a-row-in-a-latex-document
\centering
%\hspace{-25pt}
\includegraphics[trim={0cm 2cm 0cm 0.2cm},clip,width=0.66\textwidth]{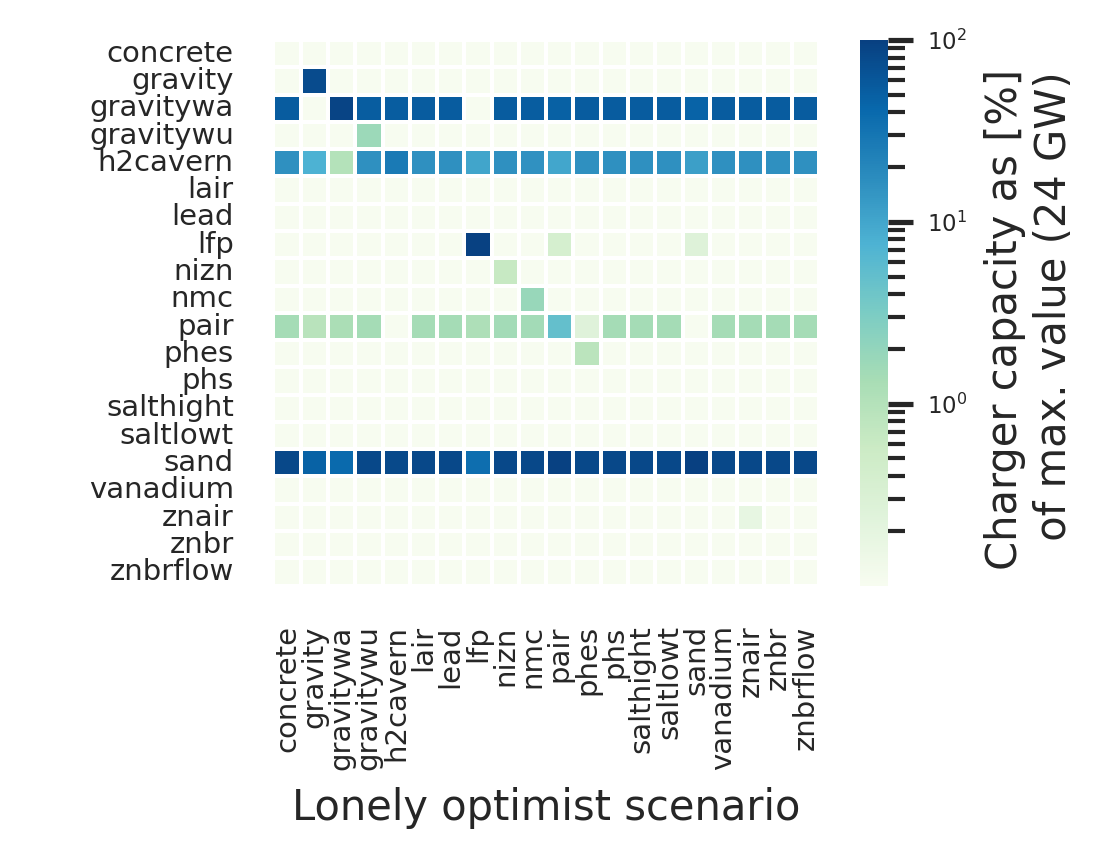}
\includegraphics[trim={0cm 2cm 0cm 0.2cm},clip,width=0.66\textwidth]{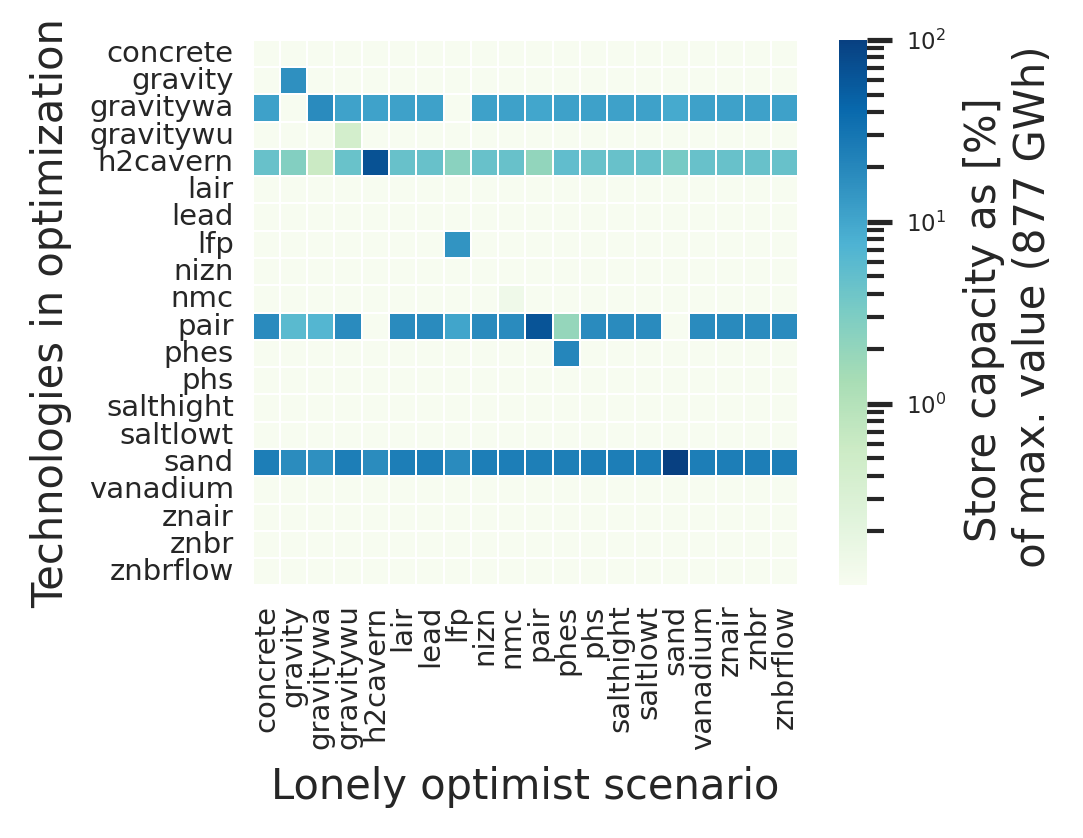}
\includegraphics[trim={0cm 0.2cm 0cm 0.2cm},clip,width=0.66\textwidth]{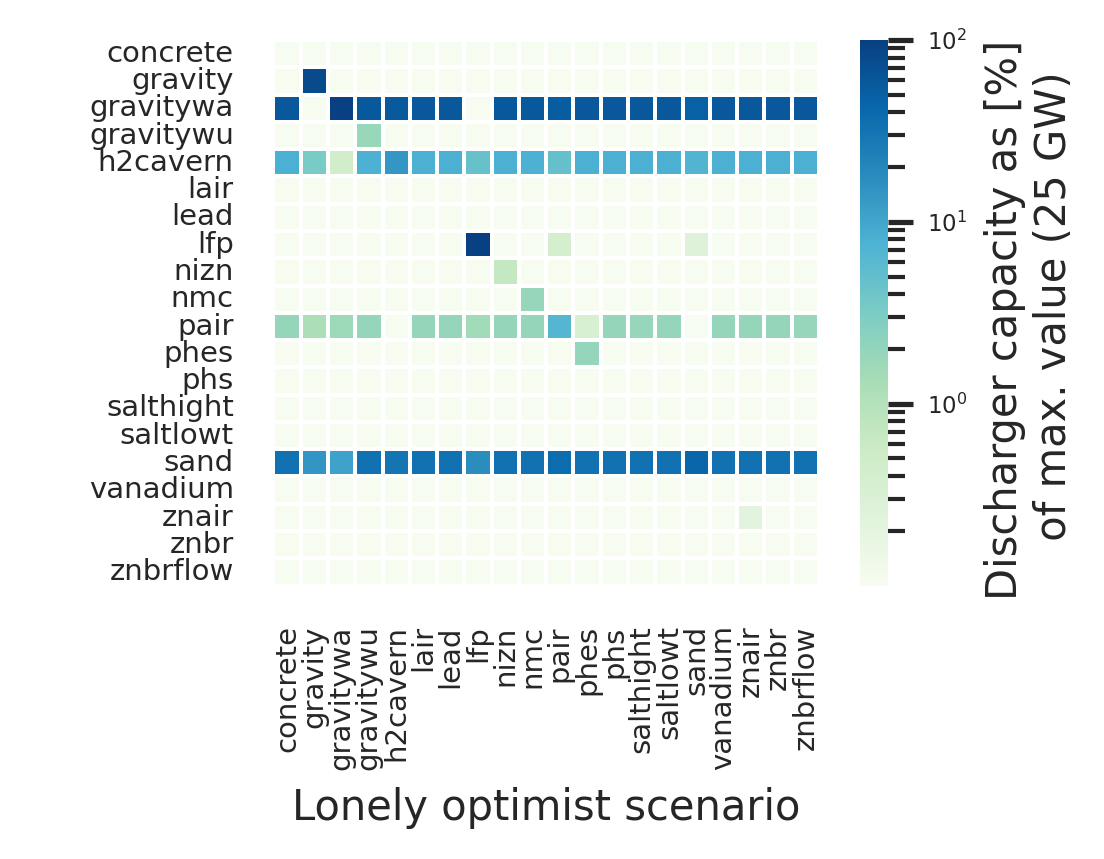}
\caption{Optimized charger, store, and discharger capacity for the lonely optimist scenario in Nigeria. All technologies on the y-axis are available for the optimization scenario in each run. One column refers thereby to one scenario run. The x-axis shows the lonely optimist scenarios, which assume optimistic capital costs assumptions (-30\%) for the mentioned technology while the others technologies on the y-axis are assumed to have pessimistic capital cost assumptions (+30\%).}\label{aueu-fig:ng-lonely-optimist-cases}
\end{figure*}

The presented results in Figure \ref{aueu-fig:ng-lonely-optimist-cases} illustrate the market potential of 20 lonely optimist scenario optimizations with 2050 techno-economic assumptions explained in Section \ref{aueu:sec:methods}. Each scenario given on the x-axis requires a single optimization run, with all technologies listed on the y-axis treated as variables. The colour gradient, normalised to the maximum value across all runs, indicates the extent to which the technologies are deployed in each optimization result. For example, the concrete lonely optimist scenario assumes optimistic capital costs for concrete-based energy storage, while the others possess pessimistic values. The following paragraphs will discuss the frequency and magnitude of storage technologies that are deployed in the optimization scenarios. These results will provide a more comprehensive understanding of the relevance and importance of each technology in the least-cost power system.

Beginning with the frequency technologies are optimized, it is observed that 9 out of 20 technologies are optimized to a significant degree, implying that not all technologies are relevant for least-cost power systems. These optimization irrelevant technologies, which are here defined as being optimized below $1\%$ of the maximally optimized technology, include concrete, lead, liquid-air, vanadium, both salt-based, pump-hydro, and any of the four zinc-based energy storage. They can likely be excluded here without consequences from further parameter studies such as global sensitivity analysis. Conversely, sand-based thermal and hydrogen cavern-based energy storage consistently provide system benefits with high certainty across all 20 scenarios examined. Further, several technologies, such as gravity-brick, underground water-based gravity, lithium ferrous phosphate (LFP), lithium nickel manganese cobalt (NMC), and pump-heat energy storage, could only compete under optimistic capital cost assumptions, while simultaneously all other technologies are attributed pessimistic values (see scenario design in Section \ref{aueu:sec:methods:scenario}). On the other hand, compressed-air, above-ground water-based gravity technologies can generally compete unless specific technologies are assumed with optimistic capital cost assumptions. For instance, the above-ground gravity storage is not optimized in a power system where gravity brick storage or Lithium LFP batteries possess optimistic capital cost assumptions. Similarly, compressed-air energy storage is not optimized when hydrogen cavern or sand energy storage is assumed to be optimistic. While analysing the frequency of energy storage optimization in various extreme parameterised scenarios is useful in determining its relevance, evaluating the magnitude is also crucial in understanding the technology's significance.

Analysing each scenario's optimised magnitude, one can observe a wide range of deployed amounts per scenario and technology. In particular, gravity and sand-based energy storage have, on average, the highest deployed amounts indicating their potentially essential role in the power system. Exploring some extremes of the scenario tree, the most optimized charger is the thermal electrode charger for the sand energy storage with $24GW$ as well as an average and minimum percentage of this value of $79\%$ and $37\%$. Similarly, the compressed air charger has a minimum, average, and maximum value of $0\%$, $1\%$ and $5\%$, while the hydrogen cavern optimization results are $1\%$, $15\%$ and $27\%$, respectively. The most optimized discharger is the lithium LFP battery inverter with $25GW$. Note that the lithium battery was not the maximum charger component due to roundtrip efficiency of $0.92\%$, which reduces the optimized amount from $25GW$ to $23GW$ such that charger and discharger are of equivalent size. Here, the relative minimum, average, and maximum values for the compressed air discharger are $0\%$, $2\%$ and $7\%$, for the hydrogen cavern-based fuel cell $1\%$, $7\%$ and $14\%$, respectively. The most optimized store is the thermal storage of the sand storage with $877GWh$ with minimum and average relative values of $16\%$ and $27\%$. Similarly, the relative minimum, average, and maximum values for the compressed air discharger are $0\%$, $16\%$ and $65\%$, for the hydrogen cavern-based fuel cell $1\%$, $7\%$ and $67\%$, respectively. As a result, while sand and gravity storage plays an important role in the power system, the other relevant technologies also sometimes contribute significantly to the least-cost power systems. 

Comparing the results to other studies, one can confirm the observation from \cite{Georgiou2020} that pumped heat energy storage can provide system benefits. However, unlike their study, liquid air energy storage is likely not optimization relevant even when optimizing multiple energy storage technologies. Interestingly, this article discovers that lithium energy storage might not be optimization relevant in many cases due to competition from other technologies, which challenges its previously overstated role in the power system decarbonisation for energy to power ratios above one hour \cite{Kittner2020, Schmidt2019}. Finally, the results reveal that gravity and sand thermal energy storage are promising technologies that warrant further investigation and inclusion in system planning.

% - While \cite{Georgiou2020} argued that both pair and phes can provide system value with an isolated system-value assessment method, we can confirm that phes can provide benefits to the system even when optimizing multiple energy storage. However, different to  \cite{Georgiou2020}, we can now say that pair is likly not optimization relevant.
% - We find that lithium energy storage can be overrated and is often not optimization relevant due to competition from other technologies
% - New to other studies, we find that especially gravity storage and sand thermal energy storage are interesting technologies to include into system planning and further to investigate.

%%% Kittner nice example
%Currently, lithium-ion battery-based energy storage remains aniche market for protection against blackouts, but our analysis shows that this could change entirely, providing flexibility and reliability for future power systems.  

\section{Technology design variation}\label{aueu:sec:designvariations}

It was found that energy storage technologies span vast sizing designs by analyzing the sizing magnitude ratios between store-to-discharger components, also known as the energy-to-power ratio (\ac{EP} ratio). One can observe EP ratios for the nine relevant storage technologies between $4-7h$ for any gravity storage, $6h$ for Lithium LFP, $8-21h$ for hydrogen cavern, $9-36h$ for sand-based, $3-19$ days for compressed air and $36$ days for pumped-heat energy storage. The results imply that for the given power system, the most critical storage categories are peak shifters (roughly $<8h$), storage that can balance mismatches also over one or multiple nights (roughly $9-36h$) and energy storage that balance through seasonal effects (roughly $7-36d$). Different to \cite{Brown2018SynergiesSystem}, the results suggest different sizing patterns for hydrogen energy storage. Compared to \ac{EP}-ratios of roughly $14-21$ days, this article finds that hydrogen storage are mostly sized to balance mismatches for one or two nights. The role of the weekly storage took the compressed-air and pumped-heat storage, which were generally not primarily optimized, reflecting that synoptic or seasonal mismatches are not as significant as predicted for the modelled power system close to the equator.

While there is an extensive design space for energy storage \cite{Sepulveda2021TheSystems, Ebbe2022}, the resulting charger-to-discharger ratios suggest that there is a general tendency that the power system benefits from oversizing the discharger components. Figure \ref{aueu-fig:foundation:storage-classification} shows that some technologies are sizing-constrained because their charger and discharger are the same components. Moreover, for sizing-unconstrained technologies such as sand-based and hydrogen cavern energy storage, the results suggest charger-to-discharger ratios between $0.28-0.61$, respectively. Only for the single case for which the pumped-heat energy storage is significantly optimized this sizing tendency is reversed such that the pumped-heat storage is sized with a charger-to-discharger ratio of $2.33$, meaning that oversizing the charger is beneficial to the power system. Similar results were found in \cite{Parzen2021BeyondSystems} for a European transmission system optimization that always oversizes, if possible, the discharger component by, on average, a factor of two or three.

% Reflect here with other papers e.g. Ebbe, Princeton, Sepuvelda, Joule paper, Strbac, Beyond cost,

% In comparison to other studies...
% - ...

%%% Kittner nice example
%Currently, lithium-ion battery-based energy storage remains aniche market for protection against blackouts, but our analysisshows that this could change entirely, providing flexibility andreliability for future power systems.  

\section{Discussion}\label{mppafrica:sec:discussion}

The importance of system-value analysis for energy storage is increasing as it allows decision-makers to evaluate the overall impact and value of multiple energy storage options within a power system. In our study, traditional system-value analysis that considers only single energy storage options in power system models may overlook significant benefits that can be obtained by designing and operating multiple energy storage options in symbiosis.

Our analysis shows that scenarios with multiple energy storage options can result in total power system cost savings of up to $3-29\%$ compared to those with only a single energy storage option. However, it is worth noting that not all energy storage options contribute equally to achieving these system benefits. Of the 20 energy storage options that were analysed, 11 are neither significantly nor frequently deployed in scenarios that considered extreme cost uncertainty, making them non-competitive.

The implications of our study for decision-makers are significant. By applying system-value analysis, investment decision-makers in industry, research, and governments can better evaluate and prioritise energy storage technologies based on their overall value in the system rather than solely on cost reduction as approached in \cite{Kittner2020}. Our findings suggest that certain energy storage technologies may not be worth investing in, while others provide good investment opportunities since they consistently provide system benefits even under high-cost uncertainty. Understanding which energy storage options provide the most frequent and significant benefits to a given power system can help decision-makers focus their limited research and deployment funds on the most viable options, potentially saving society billions of hidden costs.

Moreover, our analysis can help manufacturers and project developers design energy storage systems most valuable to a power system. Different to \cite{Sepulveda2021TheSystems, deSisternes2016TheSector, Ebbe2022}, by taking into account the benefits of multiple energy storage options, one can can derive more realistic and practical design recommendations that consider the competition and synergies between different storage options. Our study finds that energy storage technologies should be heterogeneously sized to exploit the individual system conditions and that energy-to-power ratios can vary significantly between technologies. Manufacturers can use this information to prioritise designing technologies most likely valuable to future system configurations. In contrast, project developers can apply the methods in the study to make more informed decisions about where and how to deploy energy storage systems.

However, there are limitations to our study, and it is vital to continue to improve methods, models, and data to ensure more informed decision-making in the future. Incorporating technology readiness levels, implementing realistic technology restrictions considering environmental and social limits, expanding the list of energy storage technologies, analysing various other power systems and considering competing flexibilities from other sectors such as transport and industry load shifting potentials are some areas that are more discussed in the limitation Section \ref{aueu:sec:methods:limitation}, which can be explored further. Another critical point is that technology assessments should be ideally performed globally, which requires global bottom-up model efforts such as provided in \cite{Parzen2022PyPSAEarth, osemosys-global}. Nonetheless, our study provides valuable insights into energy storage technologies. By incorporating these insights into decision-making processes, one can improve the overall system-value of energy storage and accelerate the transition to a cleaner, more sustainable and affordable energy future. To achieve this, actors in the field should avoid creating new models and instead focus on improving existing models. An open and inclusive community that promotes open research, software and data can help us progress step by step towards more informed decision-making for energy storage.

\section{Methods}\label{aueu:sec:methods}
 
\subsection{Storage data collection}
\label{aueu:sec:methods:datacollection}
% - Where is the data collected from?
% - How tech is modelled?

We extracted from a $2022$ Pacific Northwest National Laboratory (PNNL) study $20$ energy storage technologies and prepared it for reuse in any model \cite{Viswanathan_2022}. All included technologies are listed in Figure \ref{aueu-fig:foundation:storage-classification}.  The report provided techno-economic information for various storage reference sizes for $2021$ and $2030$. We focused on assumptions for the largest scale applications that range between $10-1000MW$ and $10-24h$ energy-to-power ratio. A linear extrapolation was not applicable for the $2050$ compiled data (see Table \ref{aueu-tab:2050-storage-assumptions}) as values would turn negative. To cover more of the existing non-linearity in cost developments \cite{Odenweller2022}, we created a piecewise linear approximation based on a geometric series for the years between $2034-2059$ with data points in 5-year steps.
%(see code for geometric series in supplemental list \ref{lst:geometric-series})
To explore the data, we build an interactive web application available at \href{https://pz-max-energy-storage-data-explorer-app-o5viwg.streamlit.app/}{\color{ctcolortitle}{https://pz-max-energy-storage-data-explorer-app-o5viwg.streamlit.app/}}. The original data, as well as the processing to clean and extrapolate, is integrated into an open-source tool \textit{'technology-data'} with \href{https://github.com/PyPSA/technology-data/pull/67}{\color{ctcolortitle}{https://github.com/PyPSA/technology-data/pull/67}}.

\subsection{Model and parameters}\label{aueu:sec:methods:pypsa-earth}
\begin{figure}[ht]
\centering
%\hspace{-25pt}
\includegraphics[trim={1cm 8cm 22.5cm 0cm},clip,width=0.49\textwidth]{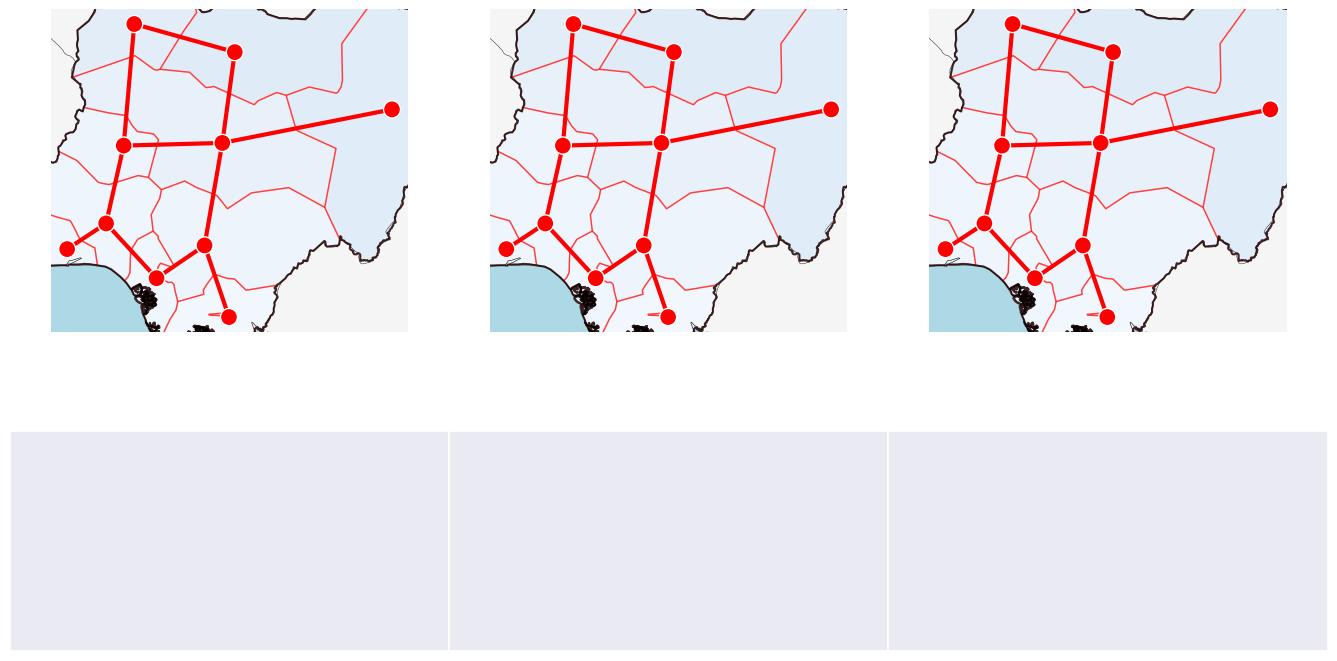}
\caption{Clustered 10 bus representation of Nigeria's power system.}
\label{aueu-fig:10-bus-ng}
\end{figure}

This study used the PyPSA-Earth model described and formulated in \cite{Parzen2022PyPSAEarth}. The model is limited to the geographical scope of the representative power system in Nigeria because its 2021 representation is already validated and described in \cite{Parzen2022PyPSAEarth}. We apply techno-economic assumptions given in Table \ref{aueu-tab:2050-system-assumptions} to represent a 2050 decarbonised power system. Further, to reduce the computational requirements, we clustered the original, open available transmission network to 10 nodes (see Figure \ref{aueu-fig:10-bus-ng}). Each node captures an area for calculating renewable potential and demand.

We contributed open-source code in \href{https://github.com/pypsa-meets-earth/pypsa-earth/pull/567}{\color{ctcolortitle}{https://github.com/pypsa-meets-earth/pypsa-earth/pull/567}} that makes adding new energy storage technologies to energy models simple. For instance, instead of adding new code for each added technology in several Python scripts, it is now possible to only add new data, and the model will automatically add these technologies. As illustrated in Figure \ref{aueu-fig:foundation:storage-classification}, energy storage with unconstrained design is modelled such that the model can independently optimize any functional component (charger, discharger, store). In contrast, design-constrained technologies such as the Lithium-battery are modelled so that the charger and discharger are constrained to be equal and share costs, representing the battery inverter. Moreover, this article excludes self-discharge losses, which were found to have negligible impact on the model outputs as our optimised technologies predominately store energy below 18 days \cite{Ebbe2022}. All required data for the energy storage technologies is described in Table \ref{aueu-tab:2050-storage-assumptions}.

\subsection{Explored scenarios}\label{aueu:sec:methods:scenario}
% ALL BELOW RUN ON THESE SCENARIOS:
% - 2030 zero emission scenario Africa
%     -- Storage Data for 2030 optimist
%     -- EQ-0.5
%     -- Weather year 2013
% ----------------------------
% - H2+battery VS all storage technologies optimized
%     -- Q&A: Do energy storage really compete? Yes, we should consider all technologies at first.
%     -- Plot: 2 x Africa map + total system costs changes
%     -- Scenario runs: 2. 
% - Parameter Sweep 1, filtering stage
%     -- Q&A: Are all energy storage relevant for the optimization? Only some, we can filter out!
%     -- Plot: Annotated heatmap with optimised capacity $https://seaborn.pydata.org/examples/spreadsheet_heatmap.html$
%     -- Scenarios runs: 20. 1 run for each of the 20 storage technologies
% - Parameter Sweep 2, global sensitivity analysis stage
%     -- Q&A: What role does uncertainty play? Important one.
%     -- Plot: Violin plot with quartile whiskers (beyond cost paper). Each Violin is market potential of one technology, one violin shows also solar, one wind capacity and one total system costs: $https://seaborn.pydata.org/examples/grouped_violinplots.html$
%     -- Scenarios runs: 1024 OR 8192. Features: charger, store, discharger capital cost per technology, VRE capital costs. Monte-Carlo require 2**N samples typically. With 4tech*3components+2VRE=14 features which results in 2**15=16384 scenarios. With 3tech*3components+2VRE=11 features which results in 2**11=2048 scenarios.
    
% TODO:
% - PREPARE PLOTS NOW
% - PREPARE CONFIGS AFTER TALKING WITH DAVIDE
Two scenario trees are explored for the $2050$ fully decarbonized power system with cost-optimal grid expansion in the model region, Nigeria. All scenarios consider a $2013$ based weather year and demand profile. The latter is scaled to align with 2050 predictions as in \cite{Parzen2022PyPSAEarth}. Era5 reanalysis data is used to derive the renewable potential calculation, and 'Shared Socioeconomic Pathways' \cite{SSP2017} are used for the 2050 hourly demand prediction (more details in \cite{Parzen2022PyPSAEarth}). Further, the scenarios include minimal variable operating and maintenance costs of $0.5 \text{\EUR{}}/MWh$ to avoid unintended storage cycling \cite{Parzen2022ReducingCosts} to reduce the risk of model distortions.

The 'single storage scenario' tree's attributes include optimising each energy storage solution in isolation, assuming $2050$ \ac{BAU} cost assumptions given in Table \ref{aueu-tab:2050-storage-assumptions}. In contrast, the 'lonely optimist scenario' tree can optimize all storage technologies simultaneously, assuming that always one storage technology is optimistic ($70\%$ of \ac{BAU} capital costs) while the others are pessimistic ($130\%$ of \ac{BAU} capital costs).

\subsection{System-value measurement}

According to Section \ref{aueu:sec:no-competition}, the concept of 'system-value' for technologies is defined as the market potential arising from the possible and probable least-cost scenarios in capacity expansion models. This definition originates from \cite{Parzen2021BeyondSystems}, which introduces the 'market potential method,' comprising two distinct components. The first component, the 'market potential indicator,' evaluates the total optimized technology size, such as an energy storage system's energy or power capacity. The second component, the 'market potential criteria,' seeks to support the decision-making process in the design of storage technologies by examining possible and probable scenarios. As per the criteria, only an optimized energy storage system can provide system-value in a least-cost power system optimization. The importance of technologies according to the system-value increases with its optimized capacity, and its provision of system-value is reinforced and more confident with its repeated optimization across multiple probable and possible scenarios. Notably, the total system costs, including any operation and investment costs, only indirectly impact the system-value assessment for technologies. Decision-makers might use it to define probable scenarios. What is likely most interesting for technology innovators, manufacturers, and regulators is the amount of a particular technology required to be deployed to benefit the power systems.

\subsection{Limitation}\label{aueu:sec:methods:limitation}
While our study provides valuable insights into the energy storage technologies that are most likely to provide system benefits, several limitations concerning the model and the data should be considered when interpreting our results and making decisions based on them.
First, our analysis assumed that all energy storage technologies have the same uncertainty range, which may be unrealistic. Incorporating a sense of technology readiness level could provide better signals on uncertainty ranges, as suggested in previous research \cite{Mankins2009}.
Second, our analysis could benefit from considering the feasibility of implementing certain energy storage technologies in all regions. For example, hydrogen cavern-based energy storage may not be feasible in some areas. Future research could include technology restrictions that account for such limitations, similar to renewable energy limitations \cite{HofmannAtlite:Series}.
Third, our study included only some possible energy storage technologies. Additional research could identify and evaluate other technologies that could potentially provide system benefits.
Fourth, our analysis did not consider competing flexibilities, such as those introduced by the transport sector or industry load-shifting potentials, which could challenge the system-value of energy storage \cite{Victoria2019TheSystem}.
Finally, to consider better uncertainty, one could consider multiple weather years and multi-year energy storage \cite{Dowling2020RoleSystems}, apply global sensitivity analysis with Monte-Carlo \cite{Mavromatidis2018} and perform near-optimal solution explorations \cite{Neumann2021TheModel}. 

\section*{Acknowledgements} 
This research was supported by UK Engineering and Physical Sciences Research Council (EPSRC) grant EP/P007805/1 and (EP/V042955/1). We would like to thank the PyPSA.org and PyPSA meets Earth teams.

\section*{Credit Authorship Contribution Statement} 

% G.F. conceptualised the study; G.F., A.V. and M.T. analysed the data and produced the figures; M.H. supervised the manuscript and acquired the funding; all authors contributed to writing and revising the manuscript.

M.P., D.F. conceptualised the study; M.P., A.K. administrated the project; M.P., D.F. contributed to the software development, M.P. performed the validation and figure production, A.K. acquired the funding; M.P. contributed to writing and revising the manuscript.

\section*{Declaration of Interests}
The authors declare no competing interests.

\newpage
\onecolumn
\section*{Supplemental material I}
\label{mppafrica:sec:supplemental material}
Energy storage and power system assumptions.
\begin{table*}[h]
\caption{Infrastructure investment cost assumptions per technology for 2050. All costs are given in real 2015 money.}
\label{aueu-tab:2050-system-assumptions}
\centering
\resizebox{0.75\textwidth}{!}{%
\begin{tabular}{lccccr}
\toprule
\textbf{\begin{tabular}[c]{@{}l@{}}Technology\\ \end{tabular}} & 
\textbf{\begin{tabular}[c]{@{}l@{}}Investment\\ (\EUR{}/kW) or\\ (\EUR{}/kWh) \end{tabular}} &
\textbf{\begin{tabular}[c]{@{}l@{}}Fixed O\&M\\(\%/year)\end{tabular}} &
\textbf{\begin{tabular}[c]{@{}l@{}}Lifetime\\(years)\end{tabular}} &
\textbf{\begin{tabular}[c]{@{}l@{}}Efficiency\\(\%)\end{tabular}} &
\textbf{\begin{tabular}[c]{@{}l@{}}Source\\ \end{tabular}}\\
\midrule
 Onshore Wind & 963 & 1.2 & 30 &   &  \cite{DEA_2019} \\ Offshore Wind & 1487 & 2.0 & 30 &   &  \cite{DEA_2019} \\ Solar PV (utility-scale) & 265 & 2.5 & 40 &   &  \cite{DEA_2019} \\ Solar PV (rooftop) & 475 & 1.6 & 40 &   &  \cite{DEA_2019} \\ Reservoir hydro & 2208 & 1.0 & 80 & 0.9 &  \cite{Schroeder_2013} \\ Run of river & 3312 & 2.0 & 80 & 0.9 &  \cite{Schroeder_2013} \\ HVDC overhead & 432 & 2.0 & 40 &   &  \cite{Hagspiel_2014} \\
  \bottomrule
\end{tabular}%
}
\end{table*}

% \clearpage
% \begin{lstlisting}[language=Python, caption={Contributed code of the required geometric series to infer realistic cost assumptions. The code is contributed to the open-source package technology-data which the model implements an energy storage data interface for.}\label{lst:geometric-series}]
% #!/usr/bin/env python
% def geo_series(nom, denom=1, no_terms=1, strt=1):
%     """
%     A geometric series is a series with a constant ratio between
%     successive terms. When moving to infinity the geometric series
%     converges to a limit.
%     https://en.wikipedia.org/wiki/Series_(mathematics)

%     Example:
%     --------
%     nom = 1  # nominator
%     denom = 2  # denominator
%     no_terms = 3
%     strt = 0  # 0 means it starts at the first term
%     result = 1/1**0 + 1/2**1 + 1/2**2 = 1 + 1/2 + 1/4 = 1.75

%     If moving to infinity the result converges to 2
%     """
%     return (sum([nom/denom**i for i in range(strt, strt+no_terms)])
% \end{lstlisting}\label{aueu-code:geometric-series}

\begin{table*}[ht!]
\caption{Electricity storage overnight investment cost assumptions per technology for 2050. Derived with geometric series applied on 2021 and 2030 PNNL data. All costs are given in real 2015 money. All costs are given in real 2015 money.}
\label{aueu-tab:2050-storage-assumptions}
\centering
\resizebox{0.75\textwidth}{!}{%
\begin{tabular}{lccccr}
\toprule
\textbf{\begin{tabular}[c]{@{}l@{}}Technology\\ \end{tabular}} & 
\textbf{\begin{tabular}[c]{@{}l@{}}Investment\\ (\EUR{}/kW) or\\ (\EUR{}/kWh) \end{tabular}} &
\textbf{\begin{tabular}[c]{@{}l@{}}Fixed O\&M\\(\%/year)\end{tabular}} &
\textbf{\begin{tabular}[c]{@{}l@{}}Lifetime\\(years)\end{tabular}} &
\textbf{\begin{tabular}[c]{@{}l@{}}Efficiency\\(\%)\end{tabular}} &
\textbf{\begin{tabular}[c]{@{}l@{}}Source\\ \end{tabular}}\\
\midrule
 Compressed-Air-Adiabatic-bicharger & 946 & 0.9 & 60 & 0.72 &  \cite{Viswanathan_2022} \\ Compressed-Air-Adiabatic-store & 5 & 0.4 & 60 &   &  \cite{Viswanathan_2022} \\ Concrete-charger & 106 & 1.1 & 35 & 0.99 &  \cite{Viswanathan_2022} \\ Concrete-discharger & 427 & 0.3 & 35 & 0.43 &  \cite{Viswanathan_2022} \\ Concrete-store & 19 & 0.3 & 35 &   &  \cite{Viswanathan_2022} \\ Gravity-Brick-bicharger & 415 & 1.5 & 41 & 0.93 &  \cite{Viswanathan_2022} \\ Gravity-Brick-store & 131 &   & 41 &   &  \cite{Viswanathan_2022} \\ Gravity-Water-Aboveground-bicharger & 365 & 1.5 & 60 & 0.9 &  \cite{Viswanathan_2022} \\ Gravity-Water-Aboveground-store & 102 &   & 60 &   &  \cite{Viswanathan_2022} \\ Gravity-Water-Underground-bicharger & 905 & 1.5 & 60 & 0.9 &  \cite{Viswanathan_2022} \\ Gravity-Water-Underground-store & 80 &   & 60 &   &  \cite{Viswanathan_2022} \\ HighT-Molten-Salt-charger & 107 & 1.1 & 35 & 0.99 &  \cite{Viswanathan_2022} \\ HighT-Molten-Salt-discharger & 428 & 0.3 & 35 & 0.44 &  \cite{Viswanathan_2022} \\ HighT-Molten-Salt-store & 78 & 0.3 & 35 &   &  \cite{Viswanathan_2022} \\ Hydrogen-charger & 190 & 0.7 & 30 & 0.7 &  \cite{Viswanathan_2022} \\ Hydrogen-discharger & 179 & 0.6 & 30 & 0.49 &  \cite{Viswanathan_2022} \\ Hydrogen-store & 4 & 0.4 & 30 &   &  \cite{Viswanathan_2022} \\ Lead-Acid-bicharger & 111 & 2.5 & 12 & 0.88 &  \cite{Viswanathan_2022} \\ Lead-Acid-store & 282 & 0.3 & 12 &   &  \cite{Viswanathan_2022} \\ Liquid-Air-charger & 451 & 0.4 & 35 & 0.99 &  \cite{Viswanathan_2022} \\ Liquid-Air-discharger & 317 & 0.5 & 35 & 0.55 &  \cite{Viswanathan_2022} \\ Liquid-Air-store & 135 & 0.3 & 35 &   &  \cite{Viswanathan_2022} \\ Lithium-Ion-LFP-bicharger & 69 & 2.2 & 16 & 0.92 &  \cite{Viswanathan_2022} \\ Lithium-Ion-LFP-store & 160 & 0.0 & 16 &   &  \cite{Viswanathan_2022} \\ Lithium-Ion-NMC-bicharger & 69 & 2.2 & 13 & 0.92 &  \cite{Viswanathan_2022} \\ Lithium-Ion-NMC-store & 182 & 0.0 & 13 &   &  \cite{Viswanathan_2022} \\ LowT-Molten-Salt-charger & 139 & 1.1 & 35 & 0.99 &  \cite{Viswanathan_2022} \\ LowT-Molten-Salt-discharger & 559 & 0.3 & 35 & 0.54 &  \cite{Viswanathan_2022} \\ LowT-Molten-Salt-store & 48 & 0.3 & 35 &   &  \cite{Viswanathan_2022} \\ Ni-Zn-bicharger & 69 & 2.2 & 15 & 0.9 &  \cite{Viswanathan_2022} \\ Ni-Zn-store & 202 & 0.2 & 15 &   &  \cite{Viswanathan_2022} \\ Pumped-Heat-charger & 723 & 0.4 & 33 & 0.99 &  \cite{Viswanathan_2022} \\ Pumped-Heat-discharger & 507 & 0.5 & 33 & 0.63 &  \cite{Viswanathan_2022} \\ Pumped-Heat-store & 7 & 0.2 & 33 &   &  \cite{Viswanathan_2022} \\ Pumped-Storage-Hydro-bicharger & 1397 & 1.0 & 60 & 0.89 &  \cite{Viswanathan_2022} \\ Pumped-Storage-Hydro-store & 57 & 0.4 & 60 &   &  \cite{Viswanathan_2022} \\ Sand-charger & 137 & 1.1 & 35 & 0.99 &  \cite{Viswanathan_2022} \\ Sand-discharger & 548 & 0.3 & 35 & 0.53 &  \cite{Viswanathan_2022} \\ Sand-store & 5 & 0.3 & 35 &   &  \cite{Viswanathan_2022} \\ Vanadium-Redox-Flow-bicharger & 111 & 2.5 & 12 & 0.81 &  \cite{Viswanathan_2022} \\ Vanadium-Redox-Flow-store & 207 & 0.2 & 12 &   &  \cite{Viswanathan_2022} \\ Zn-Air-bicharger & 129 & 2.4 & 25 & 0.79 &  \cite{Viswanathan_2022} \\ Zn-Air-store & 156 & 0.2 & 25 &   &  \cite{Viswanathan_2022} \\ Zn-Br-Flow-bicharger & 36 & 1.8 & 10 & 0.83 &  \cite{Viswanathan_2022} \\ Zn-Br-Flow-store & 357 & 0.2 & 10 &   &  \cite{Viswanathan_2022} \\ Zn-Br-Nonflow-bicharger & 129 & 2.4 & 15 & 0.89 &  \cite{Viswanathan_2022} \\ Zn-Br-Nonflow-store & 207 & 0.2 & 15 &   &  \cite{Viswanathan_2022} \\
 \bottomrule
\end{tabular}%
}
\end{table*}

\begin{table*}[ht!]
\label{aueu-table:2021-storage-assumptions}
\caption{Electricity storage overnight investment cost assumptions per technology for 2021. Derived from original PNNL data. All costs are given in real 2015 money.}
\centering
\resizebox{0.75\textwidth}{!}{%
\begin{tabular}{lccccr}
\toprule
\textbf{\begin{tabular}[c]{@{}l@{}}Technology\\ \end{tabular}} & 
\textbf{\begin{tabular}[c]{@{}l@{}}Investment\\ (\EUR{}/kW) or\\ (\EUR{}/kWh) \end{tabular}} &
\textbf{\begin{tabular}[c]{@{}l@{}}Fixed O\&M\\(\%/year)\end{tabular}} &
\textbf{\begin{tabular}[c]{@{}l@{}}Lifetime\\(years)\end{tabular}} &
\textbf{\begin{tabular}[c]{@{}l@{}}Efficiency\\(\%)\end{tabular}} &
\textbf{\begin{tabular}[c]{@{}l@{}}Source\\ \end{tabular}}\\
\midrule
Compressed-Air-Adiabatic-bicharger & 946.18 & 0.9 & 60 & 0.72 &  \cite{Viswanathan_2022} \\ Compressed-Air-Adiabatic-store & 5.448 & 0.4 & 60 &   &  \cite{Viswanathan_2022} \\ Concrete-charger & 183.635 & 1.1 & 35 & 0.99 &  \cite{Viswanathan_2022} \\ Concrete-discharger & 734.543 & 0.3 & 35 & 0.41 &  \cite{Viswanathan_2022} \\ Concrete-store & 28.893 & 0.3 & 35 &   &  \cite{Viswanathan_2022} \\ Gravity-Brick-bicharger & 415.57 & 1.5 & 41 & 0.93 &  \cite{Viswanathan_2022} \\ Gravity-Brick-store & 184.331 &   & 41 &   &  \cite{Viswanathan_2022} \\ Gravity-Water-Aboveground-bicharger & 365.63 & 1.5 & 60 & 0.9 &  \cite{Viswanathan_2022} \\ Gravity-Water-Aboveground-store & 142.417 &   & 60 &   &  \cite{Viswanathan_2022} \\ Gravity-Water-Underground-bicharger & 905.158 & 1.5 & 60 & 0.9 &  \cite{Viswanathan_2022} \\ Gravity-Water-Underground-store & 112.097 &   & 60 &   &  \cite{Viswanathan_2022} \\ HighT-Molten-Salt-charger & 183.528 & 1.1 & 35 & 0.99 &  \cite{Viswanathan_2022} \\ HighT-Molten-Salt-discharger & 734.115 & 0.3 & 35 & 0.42 &  \cite{Viswanathan_2022} \\ HighT-Molten-Salt-store & 110.714 & 0.3 & 35 &   &  \cite{Viswanathan_2022} \\ Hydrogen-charger & 1208.632 & 0.5 & 30 & 0.7 &  \cite{Viswanathan_2022} \\ Hydrogen-discharger & 1177.152 & 0.5 & 30 & 0.49 &  \cite{Viswanathan_2022} \\ Hydrogen-store & 4.779 & 0.4 & 30 &   &  \cite{Viswanathan_2022} \\ Lead-Acid-bicharger & 147.643 & 2.4 & 12 & 0.88 &  \cite{Viswanathan_2022} \\ Lead-Acid-store & 360.824 & 0.2 & 12 &   &  \cite{Viswanathan_2022} \\ Liquid-Air-charger & 500.869 & 0.4 & 35 & 0.99 &  \cite{Viswanathan_2022} \\ Liquid-Air-discharger & 351.674 & 0.5 & 35 & 0.52 &  \cite{Viswanathan_2022} \\ Liquid-Air-store & 183.974 & 0.3 & 35 &   &  \cite{Viswanathan_2022} \\ Lithium-Ion-LFP-bicharger & 94.181 & 2.1 & 16 & 0.91 &  \cite{Viswanathan_2022} \\ Lithium-Ion-LFP-store & 316.769 & 0.0 & 16 &   &  \cite{Viswanathan_2022} \\ Lithium-Ion-NMC-bicharger & 94.181 & 2.1 & 16 & 0.91 &  \cite{Viswanathan_2022} \\ Lithium-Ion-NMC-store & 361.858 & 0.0 & 16 &   &  \cite{Viswanathan_2022} \\ LowT-Molten-Salt-charger & 148.856 & 1.1 & 35 & 0.99 &  \cite{Viswanathan_2022} \\ LowT-Molten-Salt-discharger & 595.425 & 0.3 & 35 & 0.52 &  \cite{Viswanathan_2022} \\ LowT-Molten-Salt-store & 68.283 & 0.3 & 35 &   &  \cite{Viswanathan_2022} \\ Ni-Zn-bicharger & 94.181 & 2.1 & 15 & 0.89 &  \cite{Viswanathan_2022} \\ Ni-Zn-store & 337.129 & 0.2 & 15 &   &  \cite{Viswanathan_2022} \\ Pumped-Heat-charger & 802.648 & 0.4 & 30 & 0.99 &  \cite{Viswanathan_2022} \\ Pumped-Heat-discharger & 563.561 & 0.5 & 30 & 0.61 &  \cite{Viswanathan_2022} \\ Pumped-Heat-store & 29.319 & 0.1 & 30 &   &  \cite{Viswanathan_2022} \\ Pumped-Storage-Hydro-bicharger & 1397.128 & 1.0 & 60 & 0.89 &  \cite{Viswanathan_2022} \\ Pumped-Storage-Hydro-store & 57.074 & 0.4 & 60 &   &  \cite{Viswanathan_2022} \\ Sand-charger & 151.781 & 1.1 & 35 & 0.99 &  \cite{Viswanathan_2022} \\ Sand-discharger & 607.125 & 0.3 & 35 & 0.51 &  \cite{Viswanathan_2022} \\ Sand-store & 7.883 & 0.3 & 35 &   &  \cite{Viswanathan_2022} \\ Vanadium-Redox-Flow-bicharger & 147.857 & 2.4 & 12 & 0.81 &  \cite{Viswanathan_2022} \\ Vanadium-Redox-Flow-store & 311.66 & 0.2 & 12 &   &  \cite{Viswanathan_2022} \\ Zn-Air-bicharger & 129.023 & 2.4 & 25 & 0.77 &  \cite{Viswanathan_2022} \\ Zn-Air-store & 192.847 & 0.2 & 25 &   &  \cite{Viswanathan_2022} \\ Zn-Br-Flow-bicharger & 129.023 & 2.4 & 10 & 0.81 &  \cite{Viswanathan_2022} \\ Zn-Br-Flow-store & 470.192 & 0.3 & 10 &   &  \cite{Viswanathan_2022} \\ Zn-Br-Nonflow-bicharger & 129.023 & 2.4 & 15 & 0.87 &  \cite{Viswanathan_2022} \\ Zn-Br-Nonflow-store & 273.108 & 0.2 & 15 &   &  \cite{Viswanathan_2022} \\
\bottomrule
\end{tabular}%
}
\end{table*}

\begin{table*}[ht!]
\label{aueu-table:2030-storage-assumptions}
\caption{Electricity storage overnight investment cost assumptions per technology for 2030. Derived from original PNNL data. All costs are given in real 2015 money.}
\centering
\resizebox{0.75\textwidth}{!}{%
\begin{tabular}{lccccr}
\toprule
\textbf{\begin{tabular}[c]{@{}l@{}}Technology\\ \end{tabular}} & 
\textbf{\begin{tabular}[c]{@{}l@{}}Investment\\ (\EUR{}/kW) or\\ (\EUR{}/kWh) \end{tabular}} &
\textbf{\begin{tabular}[c]{@{}l@{}}Fixed O\&M\\(\%/year)\end{tabular}} &
\textbf{\begin{tabular}[c]{@{}l@{}}Lifetime\\(years)\end{tabular}} &
\textbf{\begin{tabular}[c]{@{}l@{}}Efficiency\\(\%)\end{tabular}} &
\textbf{\begin{tabular}[c]{@{}l@{}}Source\\ \end{tabular}}\\
\midrule
Compressed-Air-Adiabatic-bicharger & 946 & 0.9 & 60 & 0.72 &  \cite{Viswanathan_2022} \\ Compressed-Air-Adiabatic-store & 5 & 0.4 & 60 &   &  \cite{Viswanathan_2022} \\ Concrete-charger & 144 & 1.1 & 35 & 0.99 &  \cite{Viswanathan_2022} \\ Concrete-discharger & 576 & 0.3 & 35 & 0.43 &  \cite{Viswanathan_2022} \\ Concrete-store & 24 & 0.3 & 35 &   &  \cite{Viswanathan_2022} \\ Gravity-Brick-bicharger & 415 & 1.5 & 41 & 0.93 &  \cite{Viswanathan_2022} \\ Gravity-Brick-store & 157 &   & 41 &   &  \cite{Viswanathan_2022} \\ Gravity-Water-Aboveground-bicharger & 365 & 1.5 & 60 & 0.9 &  \cite{Viswanathan_2022} \\ Gravity-Water-Aboveground-store & 121 &   & 60 &   &  \cite{Viswanathan_2022} \\ Gravity-Water-Underground-bicharger & 905 & 1.5 & 60 & 0.9 &  \cite{Viswanathan_2022} \\ Gravity-Water-Underground-store & 95 &   & 60 &   &  \cite{Viswanathan_2022} \\ HighT-Molten-Salt-charger & 144 & 1.1 & 35 & 0.99 &  \cite{Viswanathan_2022} \\ HighT-Molten-Salt-discharger & 576 & 0.3 & 35 & 0.44 &  \cite{Viswanathan_2022} \\ HighT-Molten-Salt-store & 94 & 0.3 & 35 &   &  \cite{Viswanathan_2022} \\ Hydrogen-charger & 312 & 0.7 & 30 & 0.49 &  \cite{Viswanathan_2022} \\ Hydrogen-discharger & 414 & 0.5 & 30 & 0.7 &  \cite{Viswanathan_2022} \\ Hydrogen-store & 4 & 0.4 & 30 &   &  \cite{Viswanathan_2022} \\ Lead-Acid-bicharger & 128 & 2.4 & 12 & 0.88 &  \cite{Viswanathan_2022} \\ Lead-Acid-store & 320 & 0.2 & 12 &   &  \cite{Viswanathan_2022} \\ Liquid-Air-charger & 475 & 0.4 & 35 & 0.99 &  \cite{Viswanathan_2022} \\ Liquid-Air-discharger & 334 & 0.5 & 35 & 0.55 &  \cite{Viswanathan_2022} \\ Liquid-Air-store & 159 & 0.3 & 35 &   &  \cite{Viswanathan_2022} \\ Lithium-Ion-LFP-bicharger & 81 & 2.1 & 16 & 0.92 &  \cite{Viswanathan_2022} \\ Lithium-Ion-LFP-store & 236 & 0.0 & 16 &   &  \cite{Viswanathan_2022} \\ Lithium-Ion-NMC-bicharger & 81 & 2.1 & 13 & 0.92 &  \cite{Viswanathan_2022} \\ Lithium-Ion-NMC-store & 269 & 0.0 & 13 &   &  \cite{Viswanathan_2022} \\ LowT-Molten-Salt-charger & 144 & 1.1 & 35 & 0.99 &  \cite{Viswanathan_2022} \\ LowT-Molten-Salt-discharger & 576 & 0.3 & 35 & 0.54 &  \cite{Viswanathan_2022} \\ LowT-Molten-Salt-store & 58 & 0.3 & 35 &   &  \cite{Viswanathan_2022} \\ Ni-Zn-bicharger & 81 & 2.1 & 15 & 0.9 &  \cite{Viswanathan_2022} \\ Ni-Zn-store & 267 & 0.2 & 15 &   &  \cite{Viswanathan_2022} \\ Pumped-Heat-charger & 761 & 0.4 & 33 & 0.99 &  \cite{Viswanathan_2022} \\ Pumped-Heat-discharger & 534 & 0.5 & 33 & 0.63 &  \cite{Viswanathan_2022} \\ Pumped-Heat-store & 11 & 0.2 & 33 &   &  \cite{Viswanathan_2022} \\ Pumped-Storage-Hydro-bicharger & 1397 & 1.0 & 60 & 0.89 &  \cite{Viswanathan_2022} \\ Pumped-Storage-Hydro-store & 57 & 0.4 & 60 &   &  \cite{Viswanathan_2022} \\ Sand-charger & 144 & 1.1 & 35 & 0.99 &  \cite{Viswanathan_2022} \\ Sand-discharger & 576 & 0.3 & 35 & 0.53 &  \cite{Viswanathan_2022} \\ Sand-store & 6 & 0.3 & 35 &   &  \cite{Viswanathan_2022} \\ Vanadium-Redox-Flow-bicharger & 129 & 2.4 & 12 & 0.81 &  \cite{Viswanathan_2022} \\ Vanadium-Redox-Flow-store & 258 & 0.2 & 12 &   &  \cite{Viswanathan_2022} \\ Zn-Air-bicharger & 129 & 2.4 & 25 & 0.79 &  \cite{Viswanathan_2022} \\ Zn-Air-store & 174 & 0.2 & 25 &   &  \cite{Viswanathan_2022} \\ Zn-Br-Flow-bicharger & 81 & 2.1 & 10 & 0.83 &  \cite{Viswanathan_2022} \\ Zn-Br-Flow-store & 412 & 0.3 & 10 &   &  \cite{Viswanathan_2022} \\ Zn-Br-Nonflow-bicharger & 129 & 2.4 & 15 & 0.89 &  \cite{Viswanathan_2022} \\ Zn-Br-Nonflow-store & 239 & 0.2 & 15 &   &  \cite{Viswanathan_2022} \\
\bottomrule
\end{tabular}%
}
\end{table*}

\newpage
\clearpage
\section*{Supplemental material II}
Examples of other energy storage operations for selected scenarios.
\begin{figure*}[h]
\centering
%\hspace{-25pt}
\includegraphics[trim={0cm 0.1cm 0cm 0cm},clip,width=0.8\textwidth]{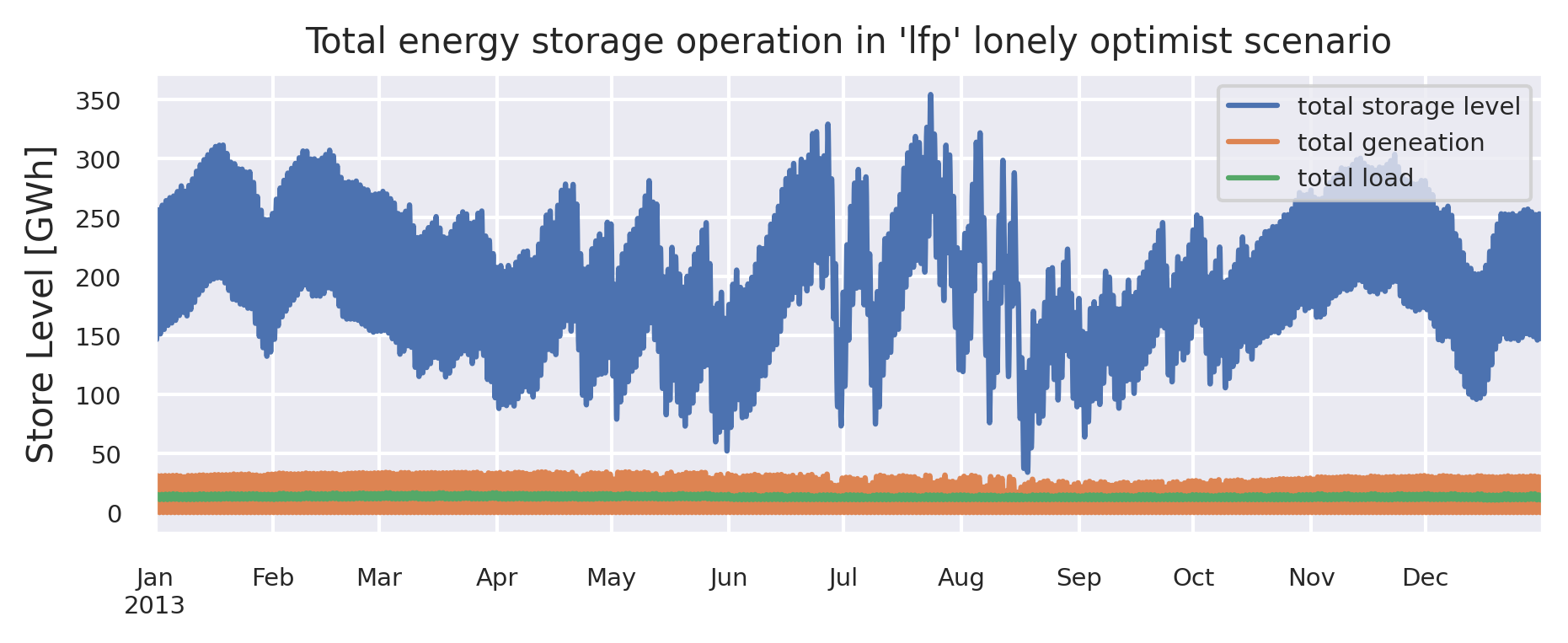}
\includegraphics[trim={0cm 0.1cm 0cm 0cm},clip,width=0.8\textwidth]{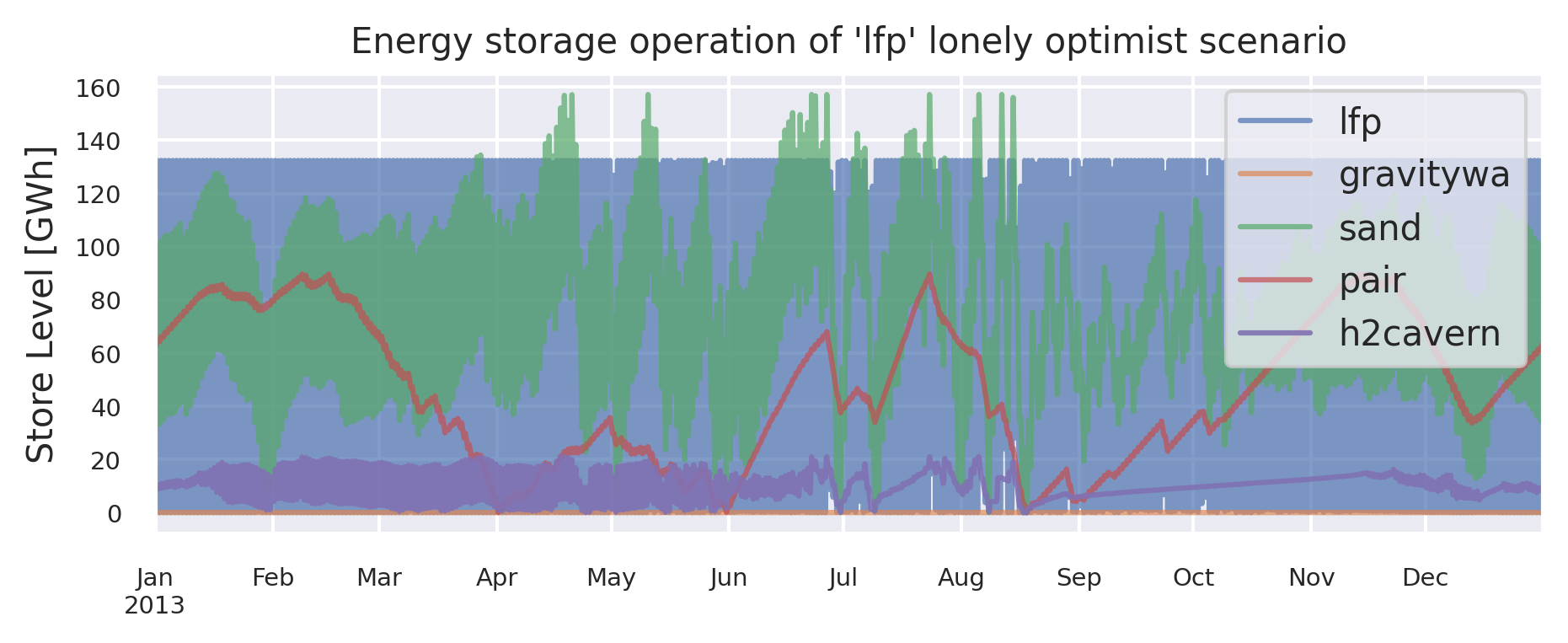}
\caption{Storage operation in lonely optimist scenario with Lithium Ferrous Phosphate (LFP) as optimistic capital costs in a group of pessimistic alternatives. The time-series is smoothed by a 12 hour rolling aggregation, where the upper figure shows the total storage operation, generation and load time-series, and the lower figure the storage operation of a selected subset of technologies. Here, the hydrogen cavern represent the blue line with maximal storage volume of around 20GWh. }
\label{aueu-fig:total-and-lfp-operation}
\end{figure*}

\clearpage
\begin{figure*}
\centering
\includegraphics[trim={0cm 0.1cm 0cm 0cm},clip,width=0.75\textwidth]{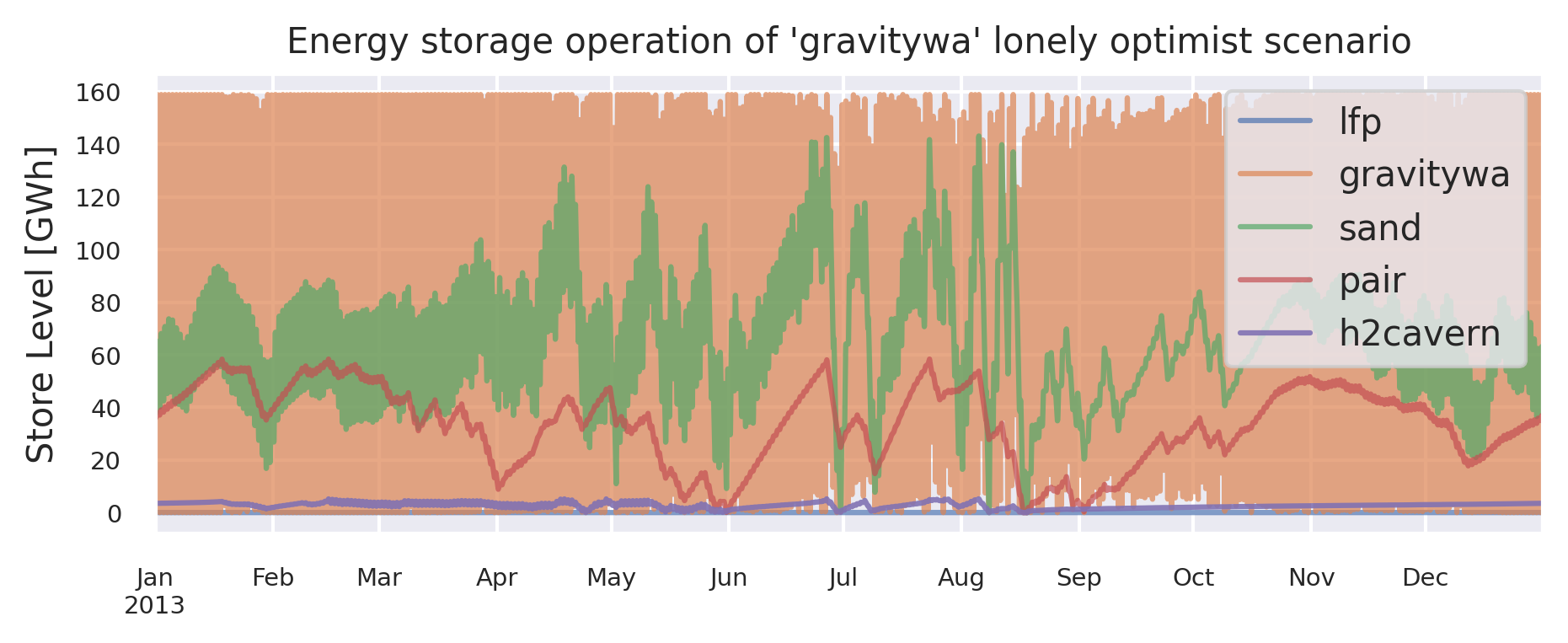}
\includegraphics[trim={0cm 0.1cm 0cm 0cm},clip,width=0.75\textwidth]{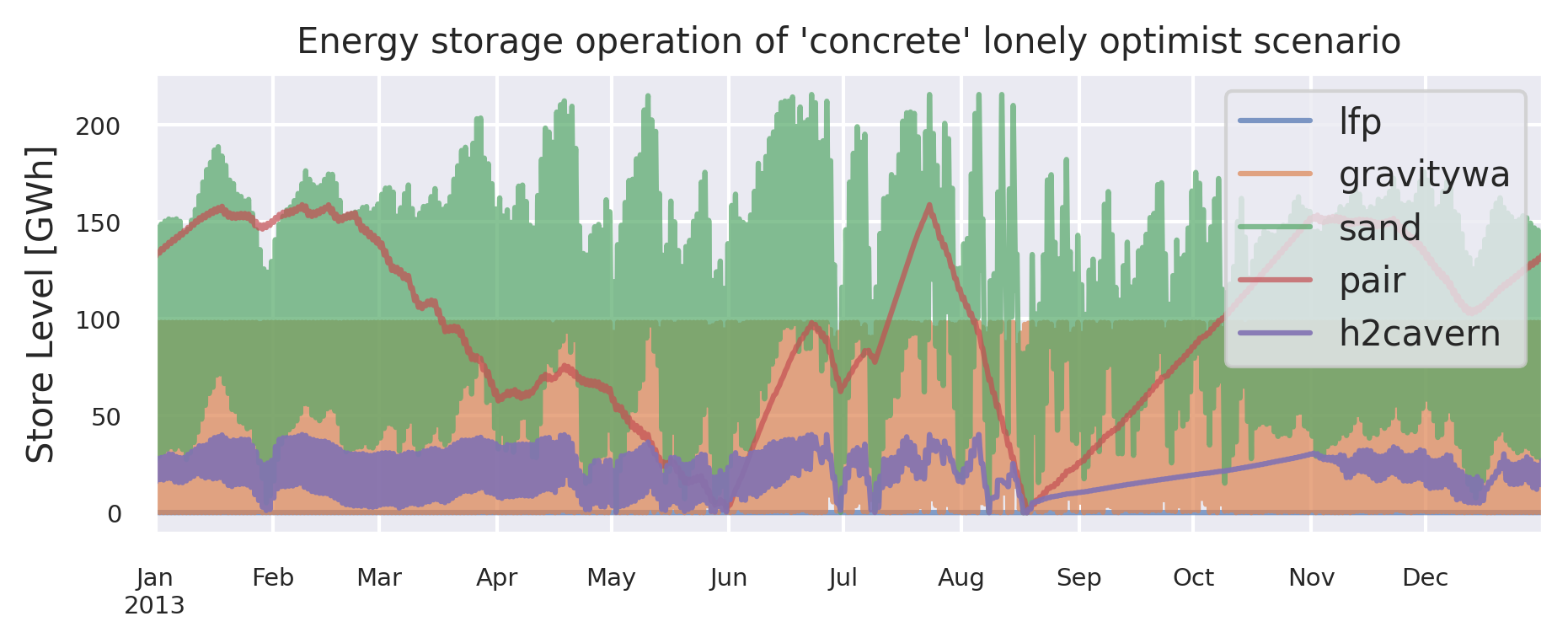}
\includegraphics[trim={0cm 0.1cm 0cm 0cm},clip,width=0.75\textwidth]{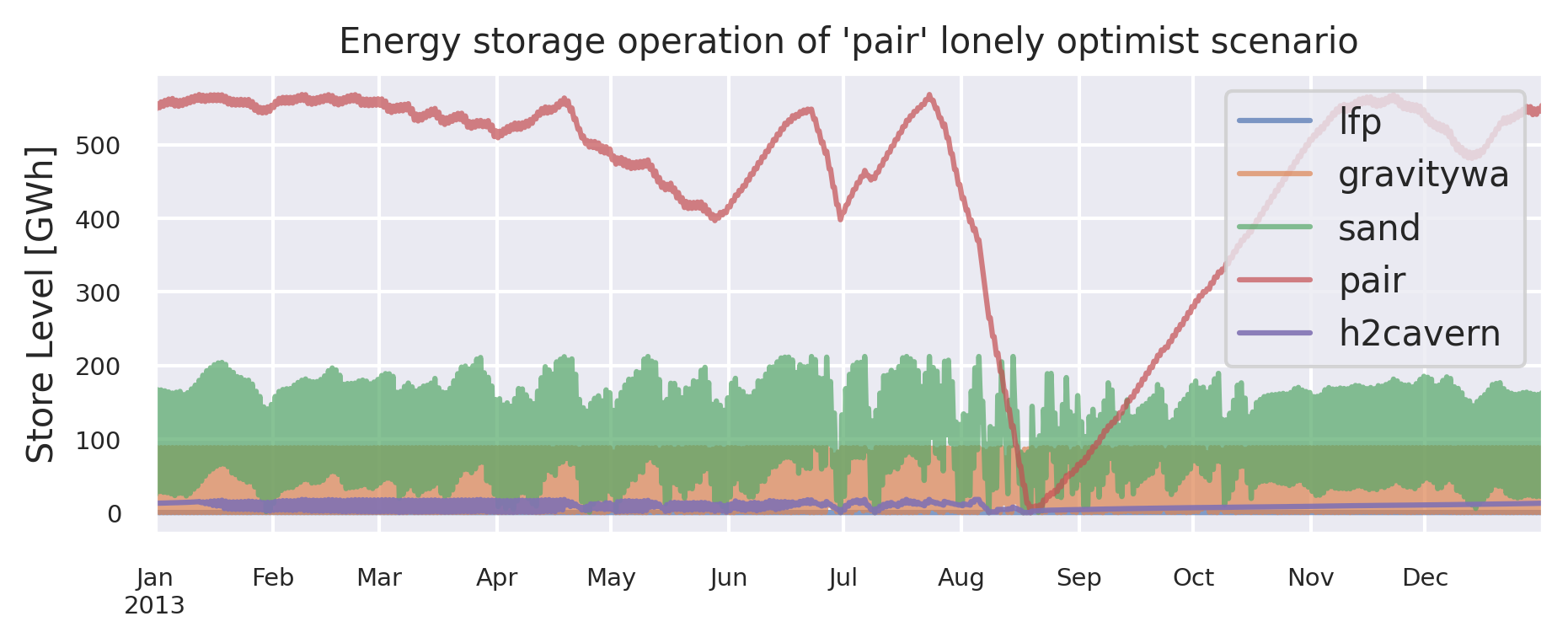}
\includegraphics[trim={0cm 0.1cm 0cm 0cm},clip,width=0.75\textwidth]{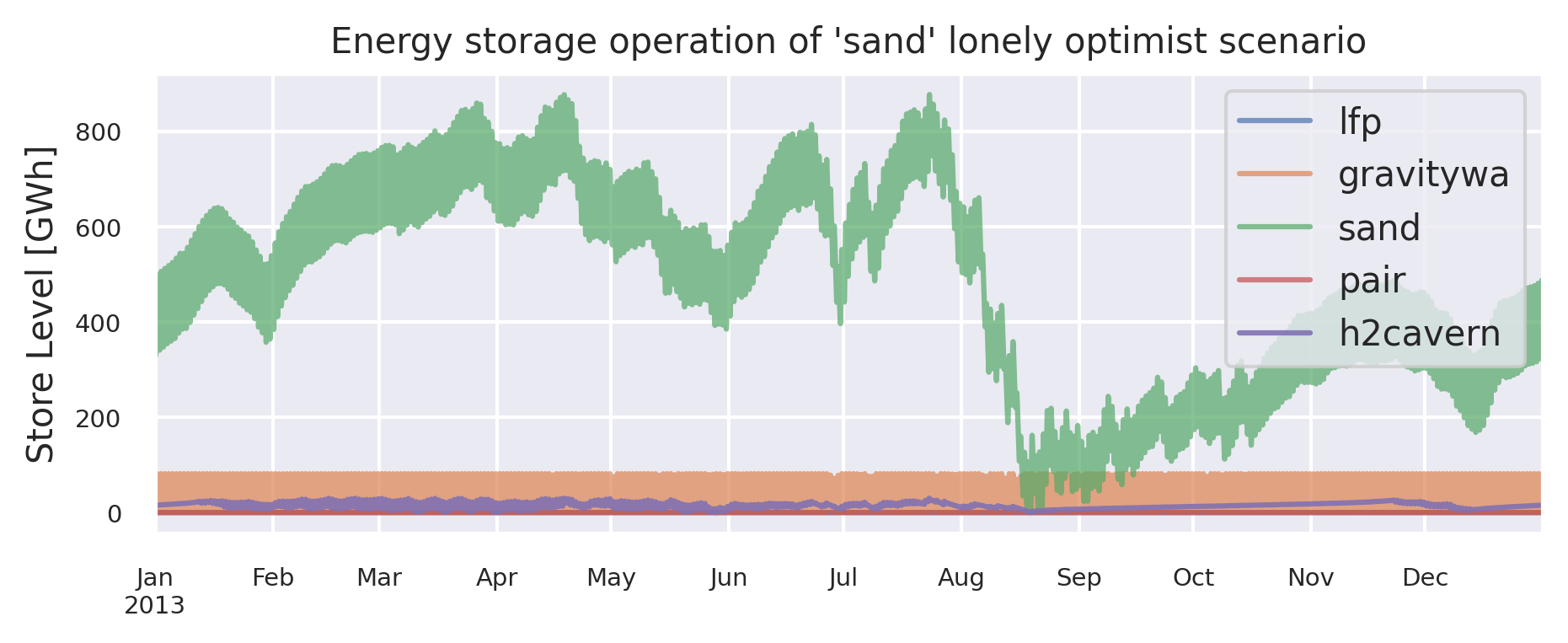}
\caption{Storage operation in lonely optimist scenario with changing optimistic storage scenarios. The time-series is smoothed by a 12 hour rolling aggregation and shows only a selected set of  technologies.}
\label{aueu-fig:gravity-concrete-pair-sand-operation}
\end{figure*}

\clearpage
\twocolumn
\printbibliography

\end{document}